\documentclass[a4paper]{article}
\usepackage{RR}
\RRNo{6253}
\usepackage{hyperref}
\usepackage{epsfig}
\usepackage{multirow}
\usepackage{amsmath}
\usepackage{amsthm}
\usepackage{pslatex}
\usepackage{pst-node}
\RRdate{July 2007}
\RRauthor{
Rumen~Andonov\thanks[irisa]{IRISA, Campus de Beaulieu, 35042 Rennes, France}%
\and
Guillaume~Collet\thanksref{irisa}%
\and
Jean-Fran\c{c}ois~Gibrat\thanks[mig]{INRA, Unit\'{e} Math\'{e}matique Informatique et G\'{e}nome UR1077, F-78352  Jouy-en-Josas}%
\and
Antoine~Marin\thanksref{mig}%
\and
Vincent~Poirriez\thanks{LAMIH, UMR CNRS 8530, Universit\'{e}  de Valenciennes,59313 Valenciennes, France}%
\and
Nikola~Yanev\thanks{University of Sofia, 5 J.~Bouchier Str., 1126 Sofia, Bulgaria}%
}
\authorhead{}
\RRtitle{Nouvelles avanc\'{e}es dans la reconnaissance de repliements}

\RRetitle{Recent Advances in Solving the Protein Threading Problem}

\RRresume{Parmi les m\'{e}thodes informatiques permettant de trouver 
la structure d'une prot\'{e}ine \`{a} partir de sa seule s\'{e}quence, 
les m\'{e}thodes par reconnaissance de repliements semblent prometteuses. 
N\'{e}anmoins, ces m\'{e}thodes demandent une tr\`{e}s grande quantit\'{e} de ressources ainsi que des algorithmes performants. 
Le logiciel FROST  (\emph{Fold Recognition Oriented Search Tool}) impl\'{e}mente une m\'{e}thode 
de reconnaissance de repliement qui est le fruit d'une \'{e}troite 
collaboration entre l'\'{e}quipe Symbiose de l'IRISA et l'unit\'{e} MIG de 
l'INRA de Jouy-en-Josas. Nous pr\'{e}sentons ici les diff\'{e}rentes 
composantes de FROST et les derni\`{e}res avanc\'{e}es r\'{e}alis\'{e}es afin 
de r\'{e}soudre efficacement le probl\`{e}me du \emph{Protein Threading}. En 
particulier, nous pr\'{e}sentons une version parall\'{e}lis\'{e}e de FROST 
permettant de r\'{e}soudre un grand nombre d'instances en un temps raisonnable.}

\RRabstract{The fold recognition methods are promissing tools for capturing the
 structure of a protein by its amino acid residues sequence but their use is 
still restricted by the needs of huge computational resources and suitable efficient 
algorithms as well.  In the recent version of  FROST  (\emph{Fold Recognition Oriented Search Tool})  
package the most efficient algorithm for solving the Protein Threading Problem (PTP) 
is implemented due to the strong collaboration between the SYMBIOSE group in IRISA and 
MIG in Jouy-en-Josas. In this paper, we present the diverse components of FROST,  emphasizing 
on the recent advances in formulating and solving new versions of the PTP and on the way of 
solving on a computer cluster a million of instances in a reasonable time. }
\RRmotcle{Reconnaissance de repliement, structure des prot\'{e}ines, parall\'{e}lisation}
\RRkeyword{Protein Threading Problem, Protein Structure, Parallel Processing}
\RRprojets{Symbiose}
\RRtheme{\THBio}

\URRennes
\begin{document}

\makeRR

\section{Introduction}
Genome sequencing projects generate an exponentially increasing amount
of raw genomic data. For a number of organisms whose genome is
sequenced, very little is experimentally known, to the point that, for
some of them, the first experimental evidence gathered is precisely
their DNA sequence. In the absence, or extreme paucity, of
experimental evidences, bioinformatic methods play a central role to
exploit the raw data. The bioinformatic process that extracts 
biological knowledge from raw data is known as annotation.

Annotation is composed of two phases:
\begin{enumerate}
\item a static phase whose purpose is to describe the basic
  ``objects'' that are found in the genome: the genes and their
  products the proteins.
\item a dynamic phase that seeks to describe the processes, i.e., the
  complex ways genes and proteins interact to create functional
  networks that underly the biological properties of the organism.
\end{enumerate}

The first phase is the cornerstone of the annotation process.  The
first step consists in finding the precise location of genes on the
chromosome.  Then, for those genes that encode proteins, the next step
is to predict the associated molecular, cellular and phenotypic
functions.  This is often referred to as \textit{in silico} functional
annotation. Different methods exist for predicting protein functions,
the most important of which are based on properties of homologous
proteins.

Homology is a key concept in biology. It refers to the fact that two
proteins are related by descent from a common ancestor. Homologous
proteins have the following properties:
\begin{itemize}
\item they may have sequences that, despite the accumulated mutations,
  still resemble the ancestor sequence;
\item their three-dimensional structures are similar to the structure
  of the ancestor;
\item they may have conserved the ancestor function, or at least a
  related function.
\end{itemize}

Therefore the principle of \textit{in silico} functional analysis,
based on homology searches, is to infer a homology relationship
between a protein whose function is known and the new protein under
study then to transfer the function of the former to the latter.

The inference of the homology relationship is based on the previously
listed properties of homologous proteins.  The first methods developed
used the first property, the conservation of the sequences, and were
based on sequence comparisons using alignment tools such as PSI-BLAST
\cite{AMS+97}.

These methods are still the workhorses of \textit{in silico} functional
annotation: they are fast and endowed with a very good statistical
criterion allowing to judge when two proteins are homologous.
Unfortunately they also have a drawback. They are very inefficient
when the proteins under study happen to be remote homologs, i.e., when
their common ancestor is very ancient.  In such a case the sequences
may have undergone many mutations and they are no longer sufficiently
similar for the proteins to be recognized as homologous.

For instance, when analyzing prokaryote genomes, these techniques
cannot provide any information about the function of a noticeable
fraction of the genome proteins (between 25\% and 50\% according to
the organism studied).  Such proteins are known as ``orphan''
proteins. One also speaks of orphan families when several homologous
proteins are found in newly sequenced genomes that cannot be linked to
any protein with a known function.

To overcome this problem new methods have been developed that are
based on the second property: the good conservation of the 3D
structure of homologous proteins. These methods are known as threading
methods, or more formally, as fold\footnote{in this context fold
  refers to the protein 3D structure} recognition methods.

The rational behind these methods is threefold:

\begin{enumerate}
\item As mentioned above, 3D structures of homologous proteins are
  much better conserved than the corresponding amino acid sequences.
  Numerous cases of proteins with similar folds and the same function
  are known, though having less than 20\% sequence identity
  \cite{brenner-1998}.

\item There is a limited, relatively small, number of protein
  structural families.  Exact figures are still a matter of debate and
  vary from 1\,000 \cite{chothia-2004} to at most a few thousands
  \cite{OJT94}. According to the statistics of the Protein Data Bank
  (PDB)\footnote{http://www.rcsb.pdb/} there are about 700 (CATH
  definition \cite{pearl-2003}) or 1\,000 (SCOP definition
  \cite{andreeva-2004}) different 3D structure families that have been
  experimentally determined so far.

\item Different types of amino acids have different preferences for
  occupying a particular structural environment (being in an
  $\alpha$-helix, in a $\beta$-sheet, being buried or exposed). These
  preferences are the basis for the empirically calculated score
  functions that measure the fitness of a given alignment between a
  sequence and a 3D structure.
\end{enumerate}

Based on these facts, threading methods consist in aligning a query
protein sequence with a set of 3D protein structures to check whether
the sequence might be compatible with one of the structures.  These
methods consist of the following components:

\begin{itemize}
\item a database of representative 3D structural templates;
\item an objective function (score function) that measures the
  fitness of the sequence for the 3D structure;
\item an algorithm for finding the optimal alignment of a sequence
  onto a structural template (with respect to the objective function);
\item a statistical analysis of the raw scores allowing the detection
  of the significant sequence-structure alignments.
\end{itemize}

To develop an effective threading method all these components must be
properly addressed. A description of the implementation of these
different components in the FROST (Fold Recognition Oriented Search
Tool) method \cite{marin-2002} is detailed in the next section. Let us
note that, from a computer scientist's viewpoint, the third component
above is the most challenging part of the treading method development.
It has been shown that, in the most general case, when variable length
alignment gaps are allowed and pairwise amino acid interactions are
considered in the score function, the problem of aligning a sequence
onto a 3D structure is NP-hard \cite{lathrop-1994}. Until recently, it
was the main obstacle to the development of efficient threading
methods. During the last few years, much progress has been
accomplished toward a solution of this problem for most real life
instances
\cite{lathrop-1996,yanev-2004,andonov-2004,xu-2003,xu-2000,balev-2004}.

Despite these improvements, threading methods, like a number of other
bioinformatic applications, have high computational requirements. For
example, in order to analyze the orphan proteins that are found in
prokaryote genomes, a back of the envelop computation shows that one
needs to align 500\,000\footnote{This figure corresponds to the number
  of sequenced genomes (500) times the average number of proteins per
  genome (3\,000) times the mean fraction of orphan proteins
  ($\frac{1}{3}$)} protein sequences with at least 1\,000 3D
structures. This represents 500 millions alignments. Solving such
quantity of alignments is, of course, not easily tractable on a single
computer.  Only a cluster of computers, or even a grid, can manage
such amount of computations. Fortunately, as we will show hereafter,
it is relatively straightforward to distribute these computations on a
cluster of processors or over a grid of computers.

Grids are emerging as a powerful tool to improve bioinformatic
applications effectiveness, particularly for protein threading. For
example, the encyclopedia of life project \cite{li-2003} integrates
123D+ threading package in its distributed pipeline. All the pipeline
processes, from DNA sequence to protein structure modeling, are
parallelized by a grid application execution environment called APST
(for Application-level scheduling Parameter Sweep Template). Another
distributed pipeline for protein structure prediction is proposed by
T. Steinke and al. \cite{steinke-2003}. Their pipeline consists in
three steps : a pre-processing phase by sequence alignments, a protein
threading phase and a final 3D refinement. Their threading algorithm
solves the alignment problem by a parallel implementation of a
Branch-\&-Bound optimizer using the score function of Xu and al.
\cite{xu-1998}. With a cluster of 16 nodes, they divided by 2 the
computation-time of aligning 572 sequences with about 37\,500
structures from the PDB.

To maintain a structural annotation database up to date (project
e-protein\footnote{http://www.e-protein.org/}), McGuffin and colleagues
describe a fold recognition method distributed on a grid with the JYDE
(Job Yield Distribution Environment) system which is a meta-scheduler
for clusters of computers. To annotate the human genome, they use
their mGen THREADER software integrated with JYDE on three different
grid systems. On these three independent clusters of 148, 243 and 192
CPUs (515 CPUs), the human genome annotations can be updated in about
24 hours.

%
%
The rest of this chapter is organized as follows. In section
\ref{sec:ori} we present basic features of the FROST method. Section
\ref{sec:vision} further details the mathematical techniques used to
tackle the difficult problem of aligning a sequence onto a 3D
structure. Section \ref{sec:distri} introduces the different
operations required in FROST to make the entire procedure modular and
describes how the modules can be distributed and executed in parallel
on a cluster of computers.  Computational benchmarks of the
parallelized version of FROST are presented in section
\ref{sec:experiments}. In section \ref{sec:future} we discuss
future research directions.
\newpage
\section{FROST: a fold recognition method}
\label{sec:ori}
  \subsection{Definition of protein cores}
  \label{subsec:cores}
Threading methods require a database of representative 3D
structures.
The Protein Data Bank (PDB) that gathers all publicly available 3D
structures contains about 40\,000 structures. However this database is
extremely redundant. Analyses of the PDB show that it contains at most
about 1\,000 different folds \cite{andreeva-2004}. In theory only
these folds need to be taken into consideration. In practice, to
obtain a denser coverage of the 3D structure space, the PDB proteins
are clustered into groups having more than 30\% sequence identity and
the best specimen of each group (in terms of quality of the 3D
structure : high resolution, small R-factor, no, or few, missing
residues) is selected. The final database contains about 4\,500 3D
structures.

For the purpose of fold recognition, the whole 3D structure is not
required, only those  parts of the structure which are characteristic of the
structural family need to be considered. This leads to the notion of
structural family core. The core is defined as those parts that are
conserved in all the 3D structures of the family and are thus
distinctive of the corresponding fold. 

There are two practical reasons for using cores:

\begin{enumerate}
\item aligning a sequence onto portions of the 3D structure that are
  not conserved is likely to introduce a noise that would make the
  detection process more difficult;
\item by definition, no insertion or deletion is permitted within core
  elements since, otherwise, they would not be conserved parts of the
  family structures.
\end{enumerate}

In protein families often one observes that the conserved framework of
the 3D structure consists of the periodic secondary structures,
$\alpha$ helices and $\beta$ strands, the loops at the surface of the
protein are variables. Accordingly, in FROST the core of the
protein structures is defined as consisting of the helices and
strands.

Hereafter we will refer to cores instead of 3D structures or folds.

  \subsection{Score function}
  \label{subsec:scores}

To evaluate the fitness of a sequence for a particular core we need an
objective (or score) function. There are two categories of score
functions:  ``local'' and ``non local''. The former ones are, in essence,
 similar to
the score functions used in sequence alignment methods. The later 
consider pairs of residues in the core and are specific of threading methods.

In threading methods, a schematic description of the core structure is
used instead of a full atomic representation.  Each residue in the
core is represented by a single \emph{site}. In FROST it is the
C$\alpha$ of the residue in the structure. Each site is characterized
by its \emph{state} which is a simplified representation of its
environment in the core. A state is defined by the type of secondary
structure ($\alpha$ helix, $\beta$ strand or coil) in which the
corresponding residue is found and by its solvent accessibility
(buried if less than 25\% of the residue surface in the core is
accessible to the solvent, exposed if more than 60\% is accessible and
intermediate otherwise). This defines 6 states, for instance the site
is located in a helix and exposed, or in a strand and buried, etc. 

In FROST we use a canonical expression for the score function.
Altschul \cite{Altschul91} has shown that the most general form of a
score for comparing sequences is a log-likelihood:

$$
score(r_{i},r_{j}) = \log\frac{P(r_{i}r_{j}|E)}{P(r_{i})P(r_{j})} 
$$

The score of replacing amino acid $r_{i}$ by amino acid $r_{j}$ is the
log of the ratio of two probabilities: 
\begin{enumerate}
\item the probability that the two amino acids are related by
  evolution, i.e., they are aligned in the sequence because they
  evolved from the same ancestral amino acid;
\item the probability that the two amino acids are aligned just by
  chance.
\end{enumerate}

If the two amino acids, on average, in a number of protein families,
are observed more often aligned than expected by chance, i.e., if the
numerator probability is greater than the product of the denominator
probabilities then the ratio is greater than 1 and the score is
positive.  On the contrary if the two amino acids are observed to be
less often aligned than expected by chance the score is negative.

These considerations led to the development of empirical substitution
matrices (for instance the PAM \cite{DSO78} or BLOSUM matrices
\cite{HH92}) that gathers the scores for replacing a given amino acid
by another one during a given period of evolution. Finding the optimal
alignment score for two sequences amounts to maximizing the
probability that these two sequences have evolved from a common
ancestor as opposed to being random sequences (assuming that the
alignment positions are independent).

Very similar matrices can be developed for threading methods, except
that we now have at our disposal an extra piece of information: the
three-dimensional structure of one of the sequences. Therefore we can
define a set of nine state-dependent substitution matrices as: 

\begin{equation}\label{loc_score}
score(R_{i},r_{j})_{S_{k}} =
\log\frac{P(R_{i}r_{j}|E)_{S_{k}}}{P(R_{i})_{S_{k}}P(r_{j})} 
\end{equation}

where $P(R_{i})_{S_{i}}$ is the probability of observing amino acid
$R_{i}$ in state $S_{k}$, $P(r_{j})$ is the background probability of
amino acid $r_{j}$ in the sequence database and
$P(R_{i}r_{j}|E)_{S_{k}}$ is the probability of observing amino acids
$R_{i}$ and $r_{j}$ aligned in sites with state $S_{k}$ in protein
families. Note that throughout this section uppercases are used for
residues that belong to the core and lower case for residues that
belong to the sequence that is aligned onto the core.

\begin{figure}[htp]

\begin{center}
\begin{small}
\begin{tabular}{ccccccccccccccc}
\textit{H}  & \textit{H} & \textit{H} & \textit{C} & \textit{C} & \textit{E}  & \textit{E} & \textit{E} & \textit{E} & -- & -- & -- & \textit{C}  & \textit{C} & \textit{C} \\
\textit{e}  & \textit{e} & \textit{e} & \textit{b} & \textit{b} & \textit{e}  & \textit{b} & \textit{b} & \textit{b} & -- & -- & -- & \textit{b}  & \textit{e} & \textit{e} \\
\textit{He}  & \textit{He} & \textit{He} & \textit{Cb} & \textit{Cb} & \textit{Ee}  & \textit{Eb} & \textit{Eb} & \textit{Eb} & -- & -- & -- & \textit{Cb}  & \textit{Ce} & \textit{Ce} \\
\textit{M}  & \textit{F} & \textit{T} & \textit{V} & \textit{N} & \textit{V}  & \textit{H} & \textit{I} & \textit{D} & -- & -- & -- & \textit{R}  & \textit{L} & \textit{Y} \\
{\bf m}  & {\bf w} & {\bf t} & -- & -- & {v} & {\bf h}  & {\bf v} & {\bf e} & {\bf h} & {\bf g} & {\bf v}  & {\bf r} & {\bf v} & {\bf y} \\
 & $\bullet$ & & & & & &$\bullet$ & & & & & & & \\
\end{tabular}
\end{small}

\begin{pspicture}(6,5.66)
\psline[linewidth=0.85pt](1,5.33)(2.33,4)(0.6,3)
\psline[linestyle=dashed,linewidth=0.85pt](.6,3)(2.33,2)(.33,1.33)(2.33,.66)(4,1.66)(5,0)(5.66,2)
\psline[linewidth=0.85pt](5.66,2)(4,3)(5.66,4)
\psline[linestyle=dashed,linewidth=0.85pt](5.66,4)(3.66,4.66)(5.66,5.33)
\psline(2.33,4)(2.66,4)
\psline(3.66,3)(4,3)
\pscircle(3,4){0.33}
\pscircle(3.33,3){0.33}
\rput(.66,5.33){{\textbf{m}}}
\rput(2,4){{\textbf{w}}}
\rput(0.33,3){{\textbf{t}}}
\rput(6,2){{\textbf{h}}}
\rput(4.33,3){{\textbf{v}}}
\rput(6,4){{\textbf{e}}}
\rput(1.66,5.33){\textit{M}}
\rput(3,4){\textit{F}}
\rput(1.66,3){\textit{T}}
\rput(2.33,4.33){\textit{$H_{e}$}}
\rput(5,2){\textit{H}}
\rput(3.33,3){\textit{I}}
\rput(4.66,4){\textit{D}}
\rput(4,3.33){\textit{$E_{b}$}}
\end{pspicture}

\end{center}

\caption{ \textbf{Upper part}: 1D alignment of two sequences the query
  sequence (5th row) is shown in bold lowercase letters, the core
  sequence (4th row) in slanted uppercase letters. The first row is the
  observed secondary structure: helix (H), strand (E) or coil (C). The
  second row is the solvant accessibility: exposed (e) or buried (b).
  The third row is the corresponding state.  Deletion are indicated by
  dashes. In the core we focus on the 2nd and 8th sites, labelled with
  black circles. The state of the 8th site is Eb, that is, an exposed
  strand. To score this position in the core we must use
  $score(I,v)_{Eb}$ the score of replacing an isoleucine by a valine
  in an exposed strand environment ($R_{i}$ = \textit{I}, $r_{j}$ =
  \textit{v} and $S_{k}$ = \textit{Eb} in the corresponding equation).
  Note also that since we are in a strand a specific gap penalty must
  be used.  \textbf{Lower part}: 3D alignment of the same two
  sequences. In the 3D structure the two above sites are in contact.
  To score this interaction we must use $score(FI,wv)_{HeEb}$ the
  score of replacing the pair \textit{FI} by the pair \textit{wv} in
  an exposed helical - buried strand environment ($R_{i}$ =
  \textit{F}, $R_{j}$ = \textit{I}, $r_{k}$ = \textit{w}, $r_{l}$ =
  \textit{v}, $S_{n}$ = \textit{He} and $S_{m}$ = \textit{Eb} in the
  corresponding equation). Here, since we are in core elements, no
  insertion/deletion is allowed.\label{fig:scores}}

\end{figure}
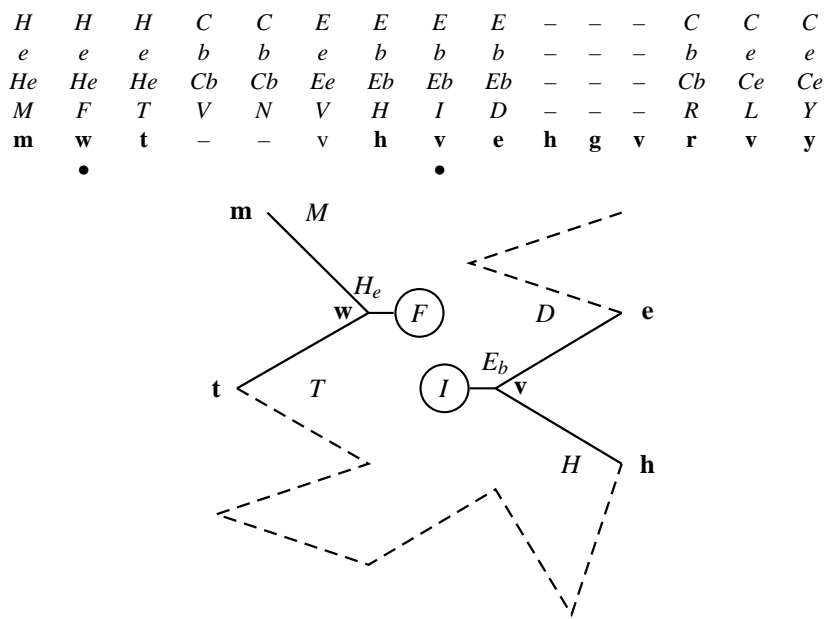

This expression represents the score for replacing amino acid $R_{i}$
by amino acid $r_{j}$ \emph{in a particular state} (see Figure
\ref{fig:scores}).  In addition, since we know the 3D structure, it is
possible to use gap penalties that prevent insertion/deletion in core
elements.  This provides a score function that is local, i.e., a score
depends on a single site in a particular sequence. However, with this
kind of score, we do not use the real 3D structure but only some of
its properties that are embodied in the state (type of secondary
structure and solvent accessibility).

In order to, explicitly, take into account the 3D structure we must
generalize these state-dependent substitution matrices. This is done
by considering pairs of residues that are \emph{in contact} in the
core. In FROST residues are defined to be in contact in a
three-dimensional structure if there exists at least one pair of
atoms, one atom from each residue side chain, for which the distance
is less than a given cut-off value. The corresponding score function
is defined as:

\begin{equation}\label{remote_score}
score(R_{i}R_{j},r_{k}r_{l})_{S_{n}S_{m}} = 
\log \frac{P(R_{i}R_{j}r_{k}r_{l}|E)_{S_{n}S_{m}}}{P(R_{i}R_{j})_{S_{n}S_{m}}P(r_{k},r_{l})}
\end{equation}

where $P(R_{i}R_{j})_{S_{n}S_{m}}$ is the probability of observing the
pair of amino acids $R_{i}$ and $R_{j}$ at sites that are in contact
in protein 3D structures and are characterized, respectively, by
states $S_{n}$ and $S_{m}$. $P(r_{k},r_{l})$ is the background
probability for the amino acid pair $r_{k}r_{l}$ in the sequence
database. $P(R_{i}R_{j}r_{k}r_{l}|E)_{S_{n}S_{m}}$ is the probability
to observe the amino acid pair $R_{i}R_{j}$ aligned with the amino
acid pair $r_{k}r_{l}$ in the structural context described by states
$S_{n}S_{m}$ in protein families.

This expression represents the score for replacing the pair of amino
acids $R_{i}R_{j}$ by the pair $r_{k}r_{l}$ in sites that are
characterized by states $S_{n}$ and $S_{m}$ and are in contact in
protein cores (see Figure \ref{fig:scores}). There are 89 such matrices.
This type of score function is non-local since it takes into account
two sites in the sequence. As we will describe in the next section the
fact that the score function is local or non-local has a profound
influence on the type of algorithm that needs to be used for aligning
the sequence onto the core.

  \subsection{Sequence-core alignment algorithms}
  \label{subsec:align} 

For local score functions there exists very efficient algorithms to
align sequences with cores. It is sufficient to borrow the algorithms
used for sequence alignments and to make the slight modifications that
are required to adapt them to our problem.  These algorithms are all
based on some forms of dynamic programming \cite{NW70, SW81} and thus
are of $O(N^{2})$, N being the size of the sequences. Besides, if the
computational requirements are of prime importance, we also have
available fast and accurate heuristics (such as BLAST and its variants
\cite{AMS+97} or FASTA \cite{Pearson00}). As shown on Figure
\ref{fig:scores} the knowledge of the 3D structure of one of the
sequence, permits the use of substitution matrices that are proper to
the state of each site in the core. Secondary structure specific gap
penalties can also be used, i.e., gap penalties that make
insertions/deletions more difficult in helices or strands. In addition
these techniques readily enable the use of sequence profiles instead
of simple sequences, a procedure that is known to improve the
sensitivity of sequence comparison methods \cite{BCH98}.

On the contrary, non-local score functions do not permit the use of
algorithms based on dynamic programming. Indeed, all dynamic
programming techniques are based on a recursive procedure whereby an
optimal solution for a given problem is built from previously found
subproblem optimal solutions.  For instance, for sequence alignments,
the optimal score for aligning two substrings $s[1..i]$ and $t[1..j]$
is obtained from the optimal solutions previously found for aligning
substrings $s[1..i-1]$ with $t[1..j-1]$, $s[1..i-1]$ with $t[1..j]$
and $s[1..i]$ with $t[1..j-1]$ by the following recurrence expression:
$$
A[i,j] = \max \left\lbrace
\begin{array}{l}
A[i-1,j] + gp \\
A[i-1,j-1] + c(s[i],t[j])\\
A[i,j-1] + gp 
\end{array}\right.
$$

where $A[k,l]$ is the optimal score for aligning substring $s[1..k]$
with substring $t[1..l]$, $gp$ is the cost of a gap and $c(s[i],t[j])$ is
the cost for aligning the $i$-th letter of string $s$ with the $j$-th
letter of string $t$.

Non-local score functions ruin this recursive procedure since, now,
the score for aligning two sequences does not exclusively depends on
the optimal score of previous subsequences but also upon interactions
with distant residues.

As a consequence, the first threading methods proposed relied on
various heuristics to align sequences onto cores, for instance Madej
et al. \cite{MGB95} used a stochastic technique in the form of a Gibbs
Monte Carlo.

Lathrop \cite{lathrop-1994} showed that, in the most general case, the
problem of aligning a sequence onto a core with a non-local score
function is NP-hard. A few years later, Akutsu and Miyano
\cite{akutsu-1999}, showed that it is MAX-SNP-hard, meaning that there
is no arbitrary close polynomial approximation algorithm, unless P =
NP.

Lathrop and Smith \cite{lathrop-1996} were the first to propose an
algorithm, based on a branch \& bound technique, that provided, for
small instances, an exact solution to the problem. Uberbacher and
colleagues \cite{xu-1998}, a couple of years later, described another
algorithm based on a divide \& conquer approach. These two algorithms
were, apparently, rather slow and only able to cope with the easiest
problems. They were not implemented in an actual threading method, to
the best of our knowledge.

At the turn of the century, new methods based on advanced mathematical
programming methods, Mixed Integer Programming (MIP), were developed
\cite{yanev-2003,xu-2003-PSB,yanev-2004,andonov-2004,balev-2004} that
were able to tackle the most difficult instances of the problem in a
reasonable amount of time. Two protein threading packages are
currently available that implement exact methods based on the latter
approach: RAPTOR\footnote{http://www.bioinformaticssolutions.com/}
\cite{xu-2003} and FROST\footnote{http://genome.jouy.inra.fr/frost/}
\cite{marin-2002}. In section \ref{sec:vision} we will describe in
more details the FROST implementation of the MIP models.  Other
interesting integer programming approaches for solving combinatorial
optimization problems that originate in molecular biology are
discussed in recent surveys \cite{lancia-2004,Blazewicz-2006}.

  \subsection{Significance of scores}
  \label{subsec:signif}

Equipped with the above techniques we are able to get an optimal score
for aligning any sequence onto a database of cores. We are now faced
with the problem of the significance of this score. Let us assume that
we have aligned a particular sequence with a core and got a score of
60. What does this score of 60 mean? Is it representative of a
sequence that is compatible with the core? In other words, if we align
a number of randomly chosen sequences with this core what kind of
score distribution are we going to obtain?  If, for a noticeable
fraction of those alignments,  one gets scores greater than or equal to
60 it is likely that the initial score is not very significant (unless
of course all the chosen sequences are related to the core).

Similar questions arise when one compares two sequences. Statistical
analyses have been carried out to study this problem and it has been
shown \cite{KA90} that the distribution of scores for ungapped local
alignments of random sequences follows an extreme value distribution.
The parameters of this distribution can be analytically calculated
from the features of the problem : type of substitution matrix used,
size of the aligned sequences, background frequencies of the amino
acids, etc. When gaped alignments are considered it is no longer
possible to perform analytical calculations but computer experiments
have shown that the shape of the empirical distribution is still an
extreme value distribution whose parameters can be readily determined
from a set of sequence comparison scores.

Such analytical calculation cannot be done for a sequence-core
alignment. In fact we do not even know the shape of the score
distribution for aligning randomly chosen sequences onto cores
although some preliminary work seems to indicate that it could also be
an extreme value distribution \cite{MFS00}.

In FROST, to solve this problem, we adopt a pragmatic, but rather
costly, approach. For each core, we randomly extract from the
database five sets of 200 sequences unrelated to the core. Each set
contains sequences whose size corresponds to a percentage of the core
size, i.e., 30\% shorter, 15\% shorter, same size as the core, 15\%
longer and 30\% longer. 
The assumption behind this procedure is that when a sequence 
is compatible with a core, its length must be similar to the 
core length ($\pm$ 30\%).
\footnote{This is the assumption in case of a global alignment. In section
\ref{sec:future} we will consider more general types of alignments: 
semi-global and local, for which  this assumption does not hold.}
We align the sequences of each set
with the core.  This provides empirical distributions of scores for
aligning sequences with different lengths onto the core.  For each
distribution we determine the median and the third quartile and we
compute a normalized score  as :

$$
S_{n} = \frac{S-q_{2}}{q_{3}-q_{2}} 
$$

where $S_{n}$ is the normalized score, S is the score of the query
sequence, $q_{2}$ and $q_{3}$ are, respectively, the median and third
quartile of the empirical distribution.

This normalized score allows us to compare the alignments of the query
sequence onto different cores. The larger the normalized score the
more probable the existence of a relationship between the sequence and
the core.  Indeed, a large normalized score indicates that the query
sequence is not likely to belong to the population of unrelated
sequences from which the score distribution was computed.
Unfortunately, since we do not know the shape nor the parameters of
the distributions, we cannot compute a precise probability for the
sequence to belong to this population of unrelated sequences.  We use
empirical results obtained on a test database to estimate when a
normalized score is significant at a 99\% level of confidence
\cite{marin-2002b,marin-2002} (see next section).

When we need to align a new query sequence whose length is not exactly
one of the above lengths that were used to pre-calculate the score
distributions, we linearly interpolate the values of the median and
the third quartile from those of the two nearest distributions. For
instance if the size of new query sequence is 20\% larger than the
size of the core, the corresponding median and third quartile values
are given by: 

$$
q^{20}_{n} = q^{15}_{n} + \frac{20 - 15}{30 - 15} (q^{30}_{n}-q^{15}_{n})
$$

where $q^{L}_{n}$ represents the median ($n=2$) or the third quartile
($n=3$) of the score distribution when sequences of length $L$ are
aligned onto the core.

  \subsection{Integrating all the components: the FROST method}
  \label{subsec:method}
FROST is intended to assess the reliability of fold assignments to a
given protein sequence (hereafter called a query sequence or query for
short) \cite{marin-2002b,marin-2002}. To perform this task, FROST used
a series of filters, each one possessing a specific scoring function
that measures the fitness of the query sequence for template cores.
The version we describe, here, possesses two filters.

The first filter is based on a fitness function whose parameters
involve only a local description of proteins and  corresponds to 
Eq. (\ref{loc_score}). 
This filter belongs to
the category of profile-profile alignment methods and is called 1D
filter. The algorithm used to find the optimal alignment score is
based on dynamic programming techniques.

The second filter employs the non local score function (\ref{remote_score}).
Because it makes use of 
spatial information, it is called a 3D filter in the following.  As
explained in section \ref{subsec:align}, this type of score function
requires dedicated algorithms for aligning the query sequence onto the
cores. The algorithm used in FROST, based on a MIP model, is further
described in the next section.

FROST functions as a sieve. The 1D filter is fast, owing to its
dynamic programming algorithm of quadratic complexity. It is used to 
compare the query sequence
with all the database cores and rank them in a list according to the normalized
scores.  Only the first $N$ cores from this list are, then,
passed to the 3D filter and aligned with the query sequence.

\begin{figure}[htp]
\begin{center}
\epsfig{figure=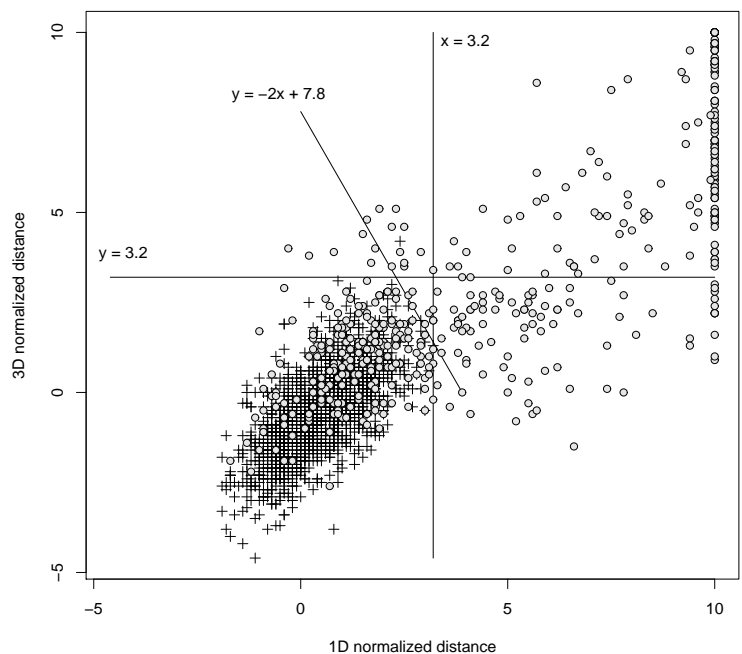, scale=.5, angle=-90}
\end{center}
\caption{Plot of the 1D score (along the x-axis) and 3D score (along
  the y-axis) for different (Q,C) pairs (where Q is a query sequence
  and C is a core). Grey open circles represent (Q,C) pairs that are
  related, black crosses (Q,C) pairs 
  that are not related, that is, respectively, the query sequence is known to have
  the same 3D structure as the core and the query sequence is known to
  have a different 3D structure from the core. The area beyond the
  lines indicated on the plot contains only 1\% black crosses, which
  are thus false positives. For this example the recall is 60\% \cite{marin-2002}.
  \label{fig:1D-3Dscores}}
\end{figure}

Each of the $N$ above cores is now characterized by two normalized
scores, one for the 1D filter and one for the 3D filter. These scores
can be plotted on a 2 dimensional diagram. As shown on Figure
\ref{fig:1D-3Dscores} this allows us to define the area on the diagram,
delimited by line equations connecting the scores, that empirically
provides a 99\% confidence threshold.

Several score functions, other than the ones described in section
\ref{subsec:scores}, can be developed. The only point that matters is
whether these functions are local or non-local. The same sieve
principle as the one described for the two above score functions is
still applicable. The difference is that now the $N$ resulting cores
are characterized by a number of scores greater than two. This makes
the visual inspection as explained above difficult and one must rely,
for instance, on a Support Vector Machine (SVM)
 algorithm to find the hyperplanes that separate
positive from negative cases.
\newpage
\section{FROST: a computer science vision}
\label{sec:vision}
  \subsection{Formal definition}
  \label{subsec:def}

In this section we give a more formal definition of protein threading
problem (PTP) and simultaneously
introduce some existing terminology. Our definition is very close to
the one given in \cite{lathrop-1996,setubal-1997}. It follows a few
basic assumptions widely adopted by the protein threading community
\cite{andonov-2004,lathrop-1998,lathrop-1996,setubal-1997,xu-2003,xu-1998}.
Consequently, the algorithms presented in the next sections can be
easily plugged in most of the existing fold recognition methods based
on threading.

\paragraph{Query sequence}
A query sequence is a string of length $N$ over
the 20-letter amino acid alphabet. This is
the amino acid sequence of a protein of unknown structure which must
be aligned with core templates from the database.

\paragraph{Core template}
All current threading methods replace the 3D coordinates of the known
structure by an abstract template description in
terms of core blocks or segments, neighbor relationships, distances,
environments, as explained in section \ref{subsec:scores}. This avoids
the computational cost of atomic-level mechanics in favor of a more
abstract, discrete representation of alignments between sequences and
cores.

\begin{figure}[htp]
\begin{center}
\psfig{figure=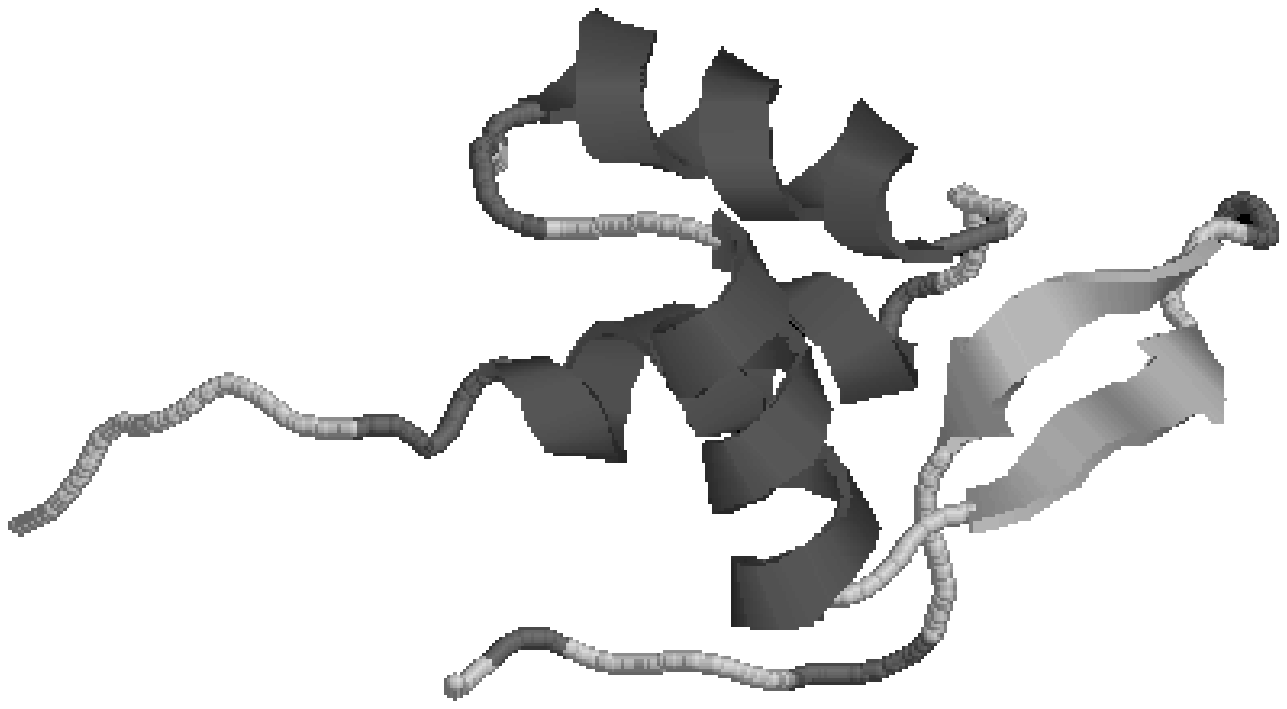, scale=.6}

(a)

\vskip24pt

\psfig{figure=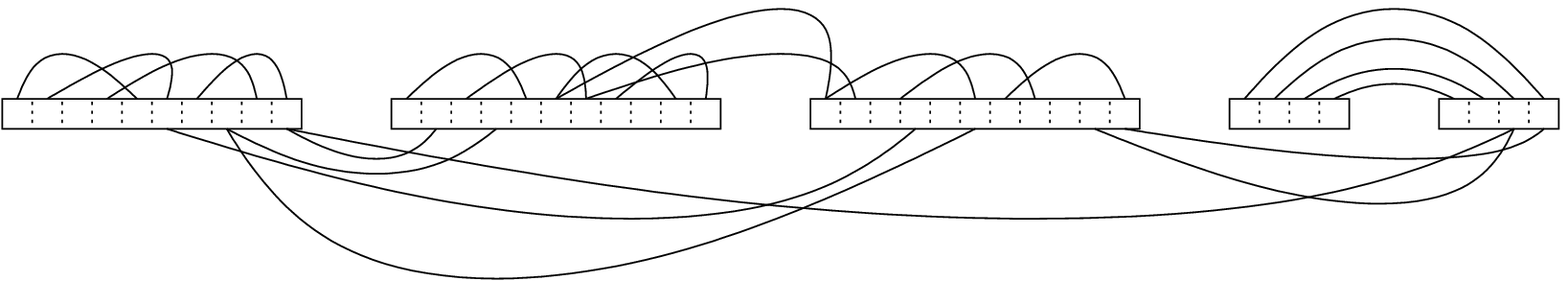, scale =.6}

(b)

\vskip24pt

\psfig{figure=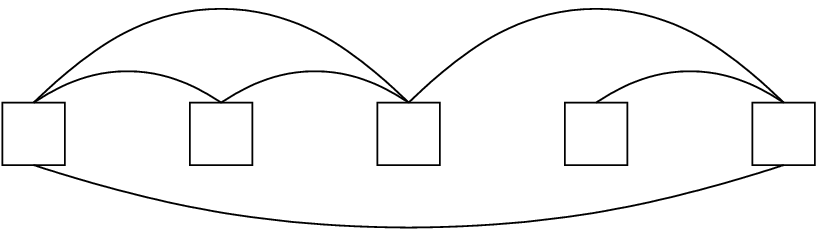, scale =.6}

(c)
\end{center}
\caption{(a) 3D structure backbone showing $\alpha$-helices,
  $\beta$-strands and loops. (b) The corresponding contact map graph.
  (c) The corresponding generalized contact map graph.}
\label{fig:3Dex}
\end{figure}

We consider that a core template is an ordered set of $m$
segments or blocks.
Segment $i$ has a fixed length of $l_i$ amino acids.
Adjacent segments are connected by variable length regions, called
loops (see Fig.~\ref{fig:3Dex}(a)).
Segments usually correspond to the most conserved parts of secondary
structure elements ($\alpha$-helices and $\beta$-strands). 
They trace the path of the conserved fold. Loops are not
considered as part of the conserved fold and consequently, the
pairwise interactions between amino acids
belonging to loops are ignored.  It is generally believed that the
contribution of such interactions is relatively insignificant. The
pairwise interactions between amino acids belonging to segments are
represented by the so-called contact map graph (see Fig.~\ref{fig:3Dex}(b)). Different
definitions for residues in contact in the core can be used, for
instance in \cite{xu-2003} they assume that two amino acids
interact if the distance between their
$C_\beta$ atoms is within $p$ \r{A} and they are at least $q$
positions apart along the template sequence (with $p=7$ and $q=4$).
There is an interaction between two segments, $i$ and $j$, if there is
at least one pairwise interaction between amino acids belonging to $i$
and amino acids belonging to $j$. Let $L \subseteq \{(i,j) \;|\; 1 \le
i < j \le m\}$ be the set of segment interactions.  The graph with
vertices $\{1,\dots,m\}$ and edges $L$ is called generalized contact
map graph (see
Fig.~\ref{fig:3Dex}(c)).

\begin{figure}[htp]
\begin{center}
\psfig{figure=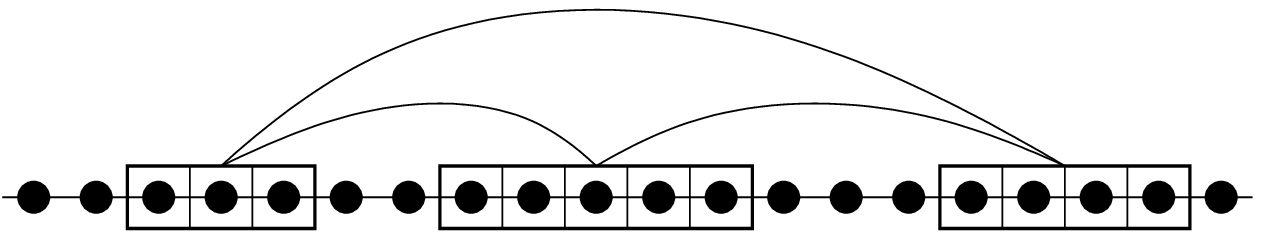, scale=.75} 

(a)

\vskip24pt

{\footnotesize
\setlength{\tabcolsep}{2pt}
\begin{tabular}{|l|rrrrrrrrrrrrrrrrrrrr|}
\hline
abs.\ position & 1 & 2 & 3 & 4 & 5 & 6 & 7 & 8 & 9 & 10 & 11 & 12 & 13 & 14 & 15 & 16 & 17 & 18 & 19 & 20\\
\hline
{ rel.\ position block 1} & { 1} & { 2} & { 3} & { 4} & { 5} & { 6} & { 7} & { 8} & { 9} & & & & & & & & & & & \\
{ rel.\ position block 2} & & & & { 1} & { 2} & { 3} & { 4} & { 5} & { 6} & { 7} & { 8} & { 9} & & & & & & & & \\
{ rel.\ position block 3} & & & & & & & & & { 1} & { 2} & { 3} & { 4} & { 5} & { 6} & { 7} & { 8} & { 9} & & & \\
\hline
\end{tabular}
}

(b)
\end{center}
\caption{(a) Example of alignment of query sequence of length 20 and
  template containing 3 segments of lengths 3, 5 and 4. (b)
  Correspondence between absolute and relative block positions.}
\label{fig:threx}
\end{figure}

\paragraph{Alignments} Let us note, first, that in this section we
adopt an inverse perspective and describe the alignment of a sequence
onto a core as positioning the segments along the sequence.  The
problem remains exactly the same but it is easier to describe this
way.  Such an alignment is called feasible if the segments preserve
their original order and do not overlap (see Fig~\ref{fig:threx}(a)).

\paragraph{}An alignment is completely determined by the starting positions of all
segments along the sequence. In fact, rather than absolute positions, it
is more convenient to use relative positions. If segment $i$ starts at the
$k$th query sequence character, its relative position is $r_i = k -
\sum_{j=1}^{i-1}l_j$. In this way the possible (relative) positions of
each segment vary between 1 and $n = N + 1 - \sum_{i=1}^m l_i$ (see
Fig.~\ref{fig:threx}(b)). The set of feasible alignments is
\begin{equation}
\mathcal{T} = \{(r_1,\dots,r_m) \;|\; 1 \le r_1 \le \dots \le r_m \le n\}.
\end{equation}
The number of possible alignments (the search space size of
PTP) is $|\mathcal{T}| =
\tbinom{m + n - 1}{m}$, which is a huge number even for small
instances (for example, if $m = 20$ and $n=100$ then $|\mathcal{T}|
\approx 2.5 \times 10^{22}$).

Most of the alignment methods impose an additional feasibility
condition, upper and lower bounds on the lengths of query zones not
covered by segments (loops). This condition can be easily incorporated
by a slight modification in the definition of relative segment
position.

In the above definition, gaps are not allowed within segments. They are confined to loops. As
explained above, the biological justification is that segments are
conserved so that the probability of insertion or deletion within them is
very small.

  \subsection{Network flow formulation}
  \label{subsec:network}

\begin{figure}[htp]

\begin{center} \epsfig{figure=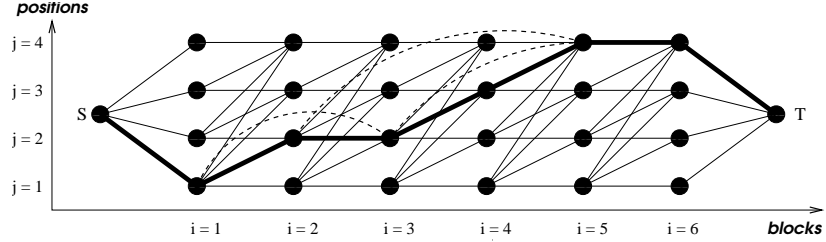, scale=.5}
\caption{Example of alignment graph. The path in thick lines corresponds to
the threading in which the positions of the blocks are
1,2,2,3,4,4. Dashed line arcs belongs to 
$A''$ where the set of segment interactions is 
$L=\{(1,3),(2,5), (3,5)\}$.\label{path}}
\end{center}
\end{figure}

This section follows the formulation proposed in \cite{yanev-2003,yanev-2004}.
In order to develop appropriate mathematical models, PTP is restated as  
a network optimization problem. Let $G(V,A)$ be a  digraph with
vertex set $V$ and arc set $A$. The vertex set $V$ is organized in
columns, corresponding to segments from the aligned core.  In each
column, each vertex correspond to a relative position of the
corresponding segment along the sequence. Then $V=\{(i,j)\; |\;
i=1,...,m,\;j=1,...,n\}$ with $m$ the number of segments and $n$ the
number of relative positions (see  Fig. \ref{path} where $m=6$ and $n=3$). A cost $C_{ij}$ is associated to each vertex $(i,j)$ as defined by the scoring function (\ref{loc_score}). 
The arc set is divided into two subsets~: $A'$
is a subset containing arcs between adjacent segments and $A''$
contains arcs between remote segments. Thus $A=A'\cup A''$ with

$$\begin{array}{l}
  A'=\{((i,j),(i+1,l))\; |\; i=1,...,m-1,\; 1 \leq j \leq l \leq n \}\\
  A''=\{((i,j),(k,l))\; |\; (i,k) \in L,\; 1 \leq j \leq l \leq n \}
\end{array}$$
To each arc $((i,j),(k,l))$ is associated a cost $D_{ijkl}$ as 
defined by the scoring function (\ref{remote_score}). The arcs from $A'$ will be referred as
 $x$-arcs and the arcs from $A''$ as $z$-arcs.

By adding two extra vertices $S$ and $T$ and the corresponding arcs
$(S,(1,k))$, $k=1,...,n$ and $((m,l),T)$, $l=1,...,n$, (considered as
$x$-arcs) one can see the one-to-one correspondence between the set of
the feasible threadings and the set of the S-T path on $x$-arcs in
$G$. We say that a S-T path \emph{activates} its vertices and
$x$-arcs. A $z$-arc is \emph{activated} by a S-T path if both ends are
on the path. We call the subgraph induced by the $x$-arcs of an S-T
path and the activated $z$-arcs \emph{augmented path}. Then PTP is
equivalent to finding the shortest augmented path in $G$.
Fig.~\ref{path} illustrates this correspondence.

  \subsection{Integer programming formulation}
  \label{subsec:integer}

\begin{figure}[htp]
 \begin{minipage}[c]{.46\linewidth}
  \centering\epsfig{figure=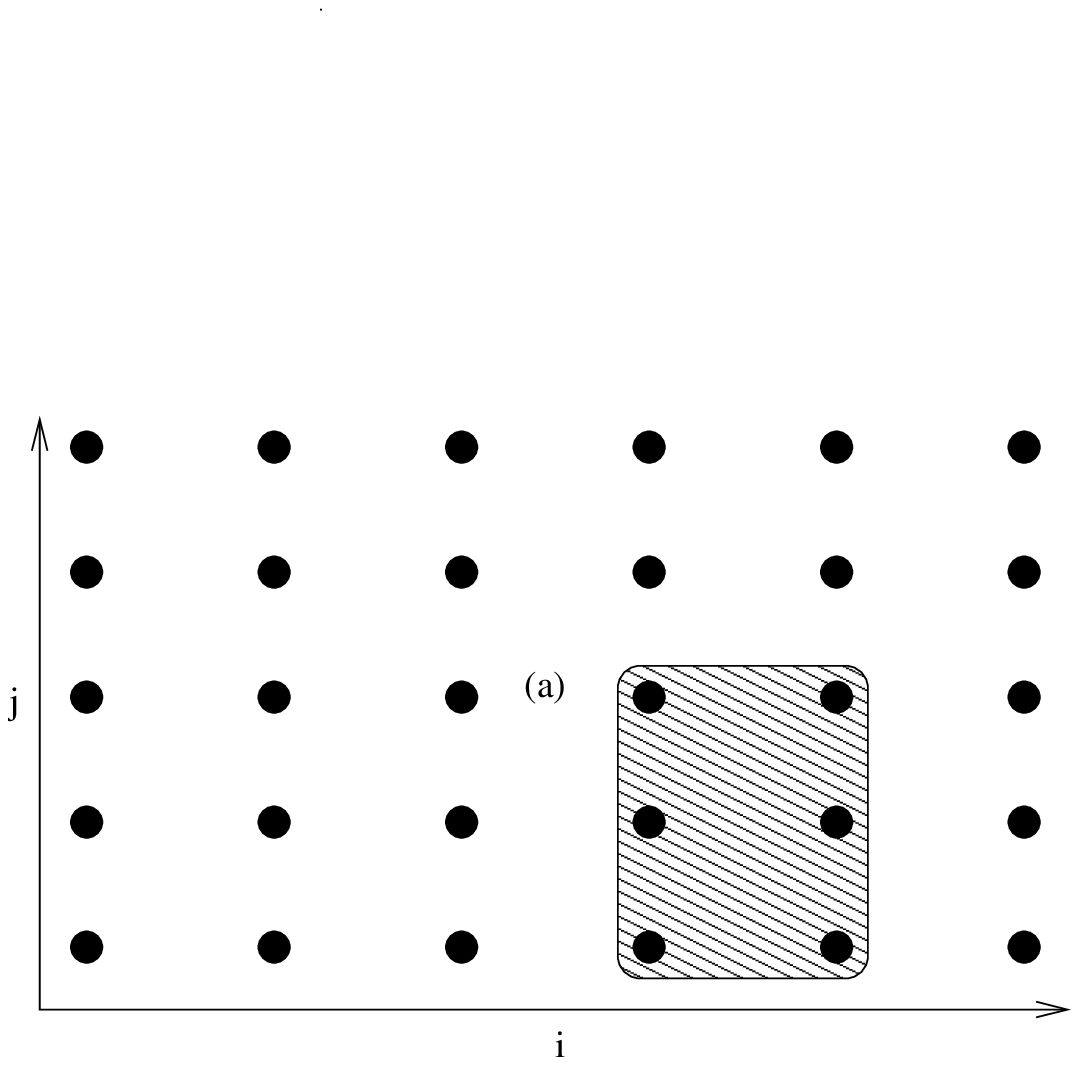, scale=.5}
 \end{minipage} \hfill
 \begin{minipage}[c]{.45\linewidth}
\setlength{\tabcolsep}{2pt}
\begin{tabular}{l}
~~\\
$y_{41} + y_{42} + y_{43} = 1$ \\
$y_{51} + y_{52} + y_{53} = 1$ \\
 $y_{41} - y_{51} \geq 0$ \\
 $y_{41} + y_{42} - y_{51} - y_{52} \geq 0$ \\
 $y_{41} + y_{42} + y_{43} - y_{51} - y_{52} - y_{53} \geq 0$ 
\end{tabular}
\end{minipage}
\caption{The effect of constraints \eqref{eq:y-assign} and \eqref{eq:y-order} on zone (a).
Exactly one vertex is activated in column four and in column five. Activating a vertex at 
position $(4,j)$ guarantees that no vertex is activated in column five below $j$.
 If a vertex is activated in  $(5,j)$, then a vertex must be activated in column four below $j$.}
\label{fig:constraint1}
\end{figure}

Let $y_{ij}$ be binary variables associated with vertices in the
previous network. Then $y_{ij}$ is one if segment $i$ is at position
$j$ and zero otherwise (vertex $(i,j)$ is activated or not). Let $Y$
be the polytope defined by the following constraints~:
\begin{align}
  &\sum_{j=1}^{n} y_{ij}  = 1 && i = 1,\dots,m \label{eq:y-assign}\\
  &\sum_{l=1}^{j} y_{il} - \sum_{l=1}^{j} y_{i+1,l} \ge 0 &&
    i = 1,\dots,m-1,\; j = 1,\dots,n-1 \label{eq:y-order}\\
    &y_{ij} \in \{0,1\} && i = 1,\dots,m,\; j = 1,\dots,n
  \end{align}
Constraint \eqref{eq:y-assign} ensures that each block is assigned to
exactly one position. Constraint \eqref{eq:y-order} describes a non-decreasing path 
in the alignment graph. These constraints are illustrated in Fig\ref{fig:constraint1}.

In order to take into account the interaction costs, we introduce a
second set of variables $z_{ijkl}\geq 0$, with $(i,k) \in L$ and $1
\le j \le l \le n$.  These variables correspond to $x$-arcs and
$z$-arcs in the network flow formulation. For the sake of readability,
we will use the notation $z_a$ for $z_{ijkl}$ with $a \in A$ the arc
set. The variable $z_{ijkl}$ is set to one if the corresponding arc is activated. 

Then, we define the following constraints~:
\begin{align}
  &y_{ij} = \sum_{l=j}^{n}z_{ijkl} && (i,k) \in L,\; j=1,...,n\label{eq:z_pred}\\
  &y_{kl} = \sum_{j=1}^{l}z_{ijkl} && (i,k) \in L,\; l=1,...,n\label{eq:z_next}\\
  &z_a \geq 0                      && a \in A
\end{align}

These constraints ensure that setting variables $y_{ij}$ and  $y_{kl}$  to one (the
path passes through these two points), activates the arc $z_{ijkl}$. 
Finding the shortest augmented path in graph $G$ (i.e. solving PTP) is
then equivalent to minimize the following function subject to the
previous constraints~:

\begin{equation}\label{obj}
\sum_{i=1}^{m} \sum_{j=1}^{n}C_{ij}y_{ij} +
\sum_{a\in A} D_a z_a
\end{equation}

This model, introduced in \cite{andonov-2004}, is known as  MYZ
model. It significantly outperforms the MIP model used in the
RAPTOR package \cite{xu-2003} for all large instances
(see \cite{andonov-2004} for more details). Both models (MYZ and RAPTOR) are solved using 
a linear programming relaxation (LP). The advantage of these 
models is that their LP relaxations give the optimal solution for 
most of the real-life instances.  They have significantly beter  performance 
than the branch \& bounds approach proposed in \cite{lathrop-1996}.
 Their drawback is their huge size 
(both number of variables and number of constraints) which makes even 
solving the LP relaxation slow. In the next section we present more 
efficient approaches for solving these models. They are based on Lagrangian relaxation.

  \subsection{Lagrangian approaches}
  \label{subsec:lagr}

Consider an integer program 
\begin{equation}
z_{IP}=\min\{ cx:x\in S\},\mbox{where } S=\{ x\in Z_{+}^{n}:Ax\geq b\}\label{ipfor}
\end{equation}
Relaxation and duality are the two main ways of determining $z_{IP}$
and upper bounds for $z_{IP}$. The linear programming relaxation is
obtained by changing the constraint $x\in Z_{+}^{n}$ in the definition
of $S$ by $x\geq0$. The Lagrangian relaxation is very convenient for
problems where the constraints can be partitioned into a set of
``simple'' ones and a set of ``complicated'' ones. Let us assume for
example that the complicated constraints are given by $A^{1}x\geq
b^{1}$, where $A^{1}$ is $m\times n$ matrix, while the simple
constraints are given by $A^{2}x\geq b^{2}$. Then for any $\lambda\in
R_{+}^{m}$ the problem
$$z_{LR}(\lambda)=\min_{x\in Q}\{ cx+\lambda(b^{1}-A^{1}x)\}$$
where $Q=\{ x\in Z_{+}^{n}:A^{2}x\geq b^{2}\}$ is the Lagrangian
relaxation of (\ref{ipfor}), i.e. $z_{LR}(\lambda)\leq z_{IP}$ for
each $\lambda\geq 0$. The best bound can be obtained by solving the
Lagrangian dual $\displaystyle z_{LD}=\max_{\lambda\geq0}z_{LR}(\lambda)$.
It is well known that the relation $z_{IP}\geq z_{LD}\geq z_{LP}$ holds.

  \subsection{Lagrangian relaxation}
  \label{subsec:relax}

\begin{figure}[htp]
\begin{minipage}[t]{.45\textwidth}
\begin{center}
\epsfig{figure=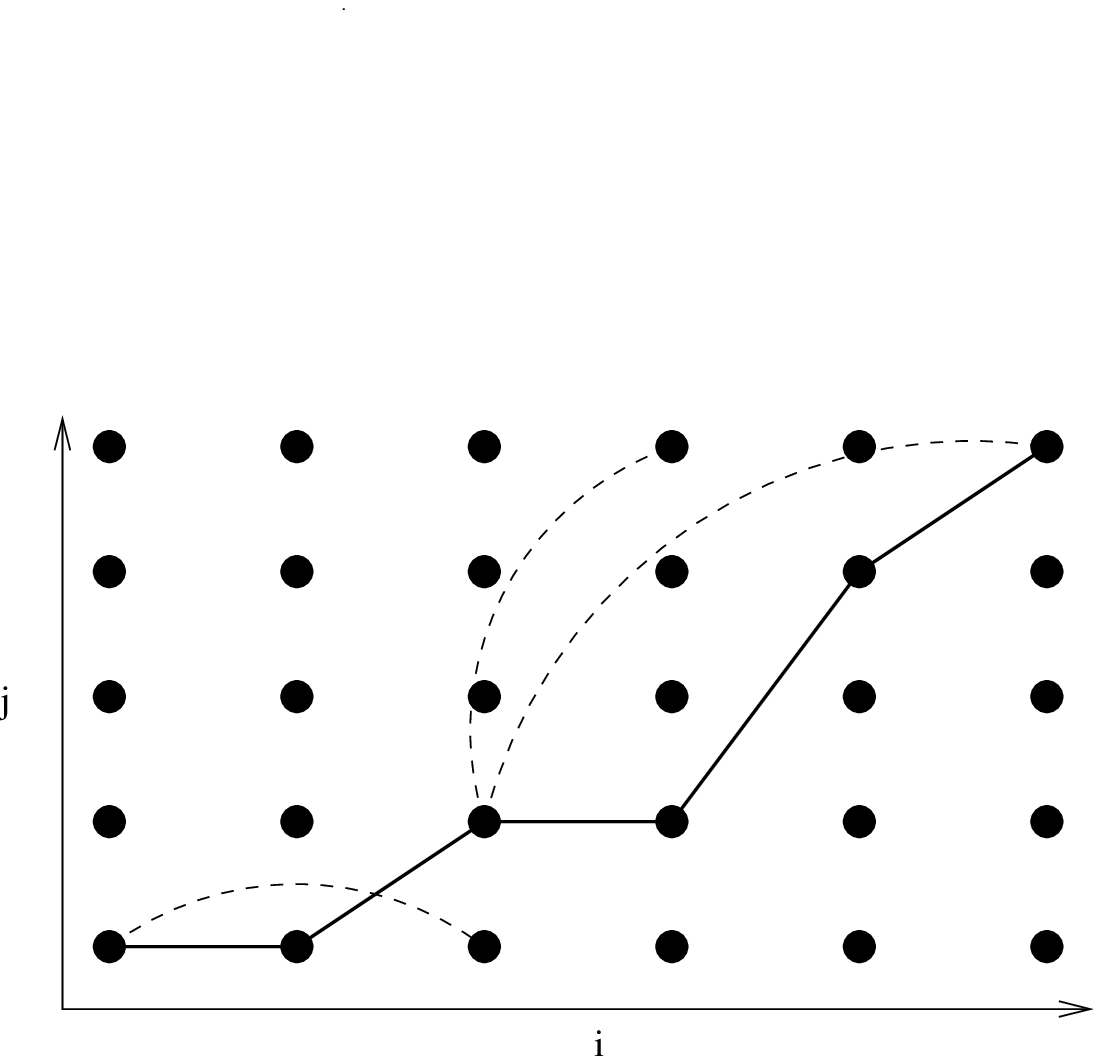, scale=.5}
(a)
\end{center}
\end{minipage}
\hfill
\hfill
\begin{minipage}[t]{.45\textwidth}
\begin{center}
\epsfig{figure=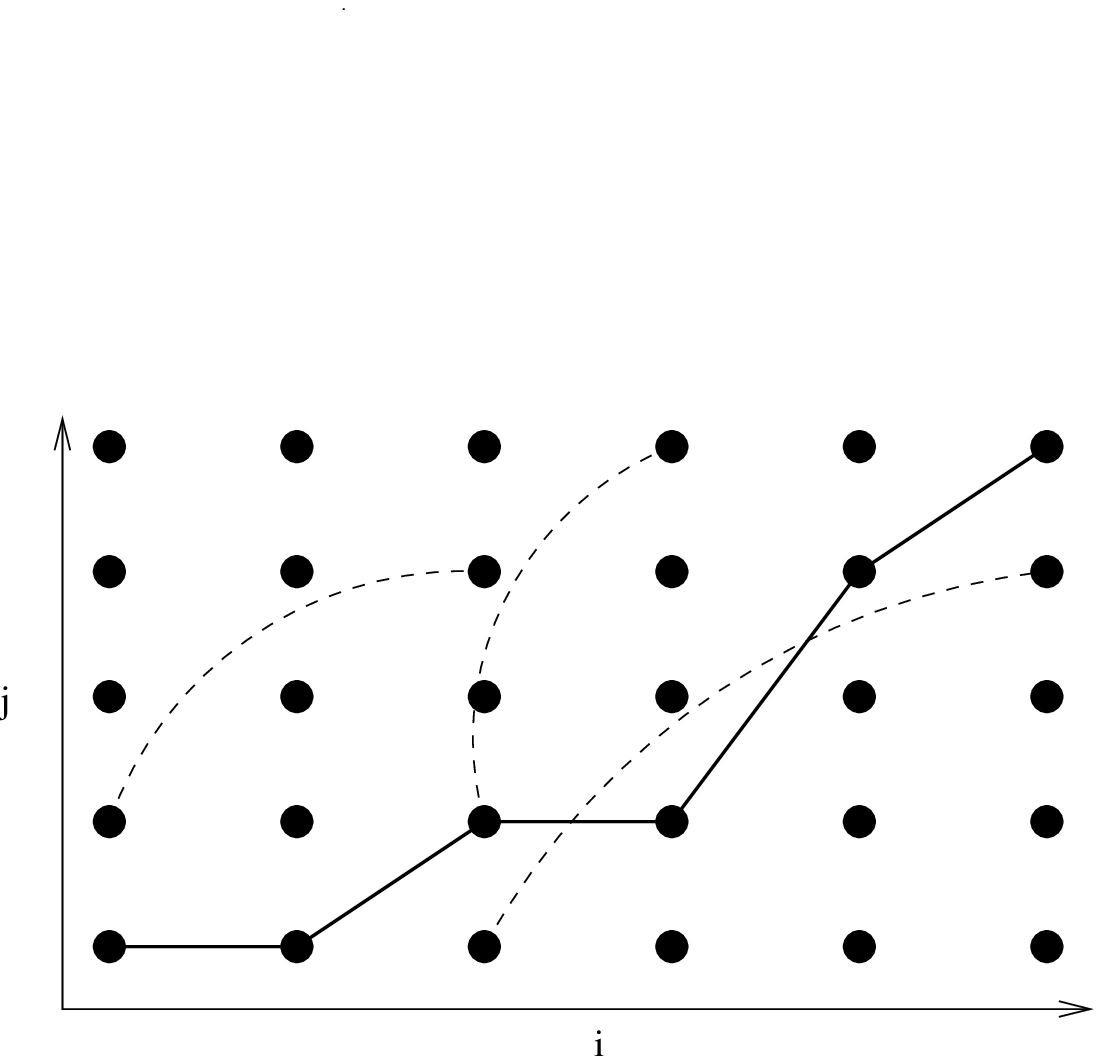, scale=.5}
(b)
\end{center}
\end{minipage}
\caption{Example of a threading instance with $m=6$ blocks and $n=5$
  free positions. The set of segment interactions is
  $L=\{(1,3),(3,4),(3,6)\}$. (a) The Lagrangian relaxation sets the
  right end of any arc free. The solution for the relaxed
  problem could not satisfy the original constraints. (b) The
  Lagrangian relaxation sets both right and left ends of arcs free.}
\label{fig:relax}
\end{figure}

We show now how to apply Lagrangian relaxation (LR) taking Eq.
\eqref{eq:z_next} as a complicated constraint. Recall that this 
constraint insures that the $y$-variables and the $z$-variables select
the same position of segment $k$. By relaxing such a constraint, we
relax the right end of a z-arcs. This means that an arc can be activated
even though its right end is not on the path, as it is illustrated  in
Fig\ref{fig:relax}(a).  For a fixed $\lambda$, 
the relaxed augmented path problem  obtained in this way can be solved  
in a polynomial time using a dynamic programming (see \cite{yanev-2006}). 

In order to find the Lagrangian dual  $z_{LD}$ one has
to look for the maximum of a concave piecewise linear function. This
appeals for using the so called  sub-gradient optimization technique. For
the function $z_{LR}(\lambda)$, the vector $s^{t}=b^{1}-A^{1}x^{t}$,
where $x^{t}$ is an optimal solution to ${\min_{x}\{ cx+\lambda^{t}(b^{1}-A^{1}x)\}}$,
is a sub-gradient at $\lambda^{t}$. 
The following sub-gradient algorithm is an analog of the steepest ascent method for maximizing a function:
\begin{itemize}
\item (Initialization): Choose a starting point $\lambda^{0}$, $\Theta_{0}$
and $\rho$. Set $t=0$ and find a sub-gradient $s^{t}$. 
\item While $s^{t}\neq0$ and $t<t_{\max}$ do \{ $\lambda^{t+1}=\lambda^{t}+\Theta_{t}s^{t};\Theta_{t+1}= 
\rho \Theta_t$, $t\leftarrow t+1$; find $s^{t}$\}
\end{itemize}
This algorithm stops either when $s^{t}=0$, (in which case $\lambda^{t}$
is an optimal solution) or after a fixed number of iterations  $t_{\max}$. 
The parameter $0 <\rho < 1$ determines the decrease of the 
sub-gradient step.

Note that for each $\lambda$ the solution defined by the $y$-variables
is feasible for the original problem. In this way at each iteration of
the sub-gradient optimization we have a heuristic solution. At the end
of the optimization we have both lower and upper bounds on the optimal
objective value.

Symmetrically, we can relax the left end of each link or even relax
the left end of one part of the links and the right end of the rest
(see figure \ref{fig:relax}(b)).  This approach is  used in
\cite{balev-2004}. The same paper describes a branch-and-bound
algorithm using this Lagrangian relaxation instead of the LP
relaxation.  This is the  default algorithm in the FROST package.

\begin{figure}[htp]
\begin{center}
\epsfig{figure=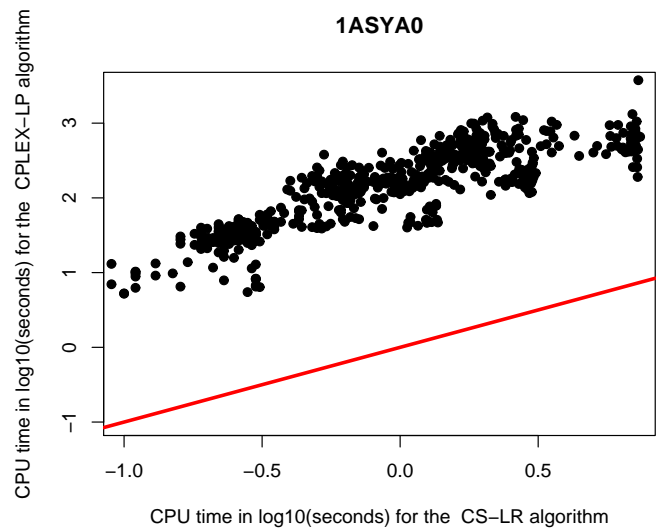, scale=.60}
\caption{Cost-Splitting Relaxation versus LP Relaxation\label{times1}.
  Plot of times in seconds with the CS algorithm on the $x$-axis and
  the LP algorithm from \cite{andonov-2004} on the $y$-axis. Both
  algorithms compute approximate solutions for 962 threading instances
  associated with the template 1ASYA0 from the FROST core database.
  The line $y=x$ is shown on the plot. A significant performance gap
  is observed between the algorithms. For example point
  $(x,y)=(0.5,3)$ corresponds to a case where CS is $10^{2.5}$ times
  faster than LP relaxation.  These results were obtained on an
  Intel(R) Xeon(TM) CPU 2.4~GHz, 2 GB RAM, RedHat 9 Linux. The MIP
  models were solved using CPLEX 7.1 solver (see \cite{veber-2005} for
  more details).  }
\end{center}
\end{figure}

Another relaxation, called \emph{cost-splitting} (CS), 
is presented in \cite{veber-2005}. The results presented in this paper 
clearly show that CS slightly outperforms LR, and both (LR and CS) relaxations  
are significantly faster than LP (see Fig.~\ref{times1}). The
interested reader can find further details concerning 
these approaches in \cite{yanev-2006}.

\newpage
\section{Dividing FROST into modules for distribution over a cluster} 
\label{sec:distri}
The following two sections are based on the results presented in \cite{poirriez-2005}. 
  \subsection{Amount of computation to be done}
In section \ref{subsec:method} we described the FROST functioning. From a
computational viewpoint, this procedure can be divided into 2 phases:
the first one is the computation of score distributions (hereafter
called phase $D$) and the second one is the alignment of the sequence
of interest with the dataset of templates (hereafter called phase $E$
for evaluation) making use of the previously calculated distributions.
These two phases are repeated for each filter (1D and 3D). We denote
by \verb|Ali1D(Q,C)| the process of aligning a query sequence (Q) with
a core (C) in the 1D filter and by \verb|Ali3D(Q,C)| the more computer
intensive alignment process of the 3D filter. Although we have a very
efficient implementation of the corresponding algorithm based on a
Lagrangian relaxation technique, computing the score distributions for
all the templates takes more than a month when performed sequentially.\\

The whole procedure requires the following computations:
\begin{enumerate}
\item Phase $D$: align non homologous sequences in order to obtain the
  scores distributions for all templates and all filters. Since five
  distributions are associated to any template, and there are about
  200 sequences for each distribution, this procedure needs solving
  about 1,200,000 quadratic problems \verb+Ali1D+ and the same amount
  of NP-complete problems \verb+Ali3D+.

\item Phase $E$: align the query with the dataset of templates which
  requires solving several hundreds of quadratic problems \verb+Ali1D+
  and $N$ NP-complete problems \verb+Ali3D+ (where $N$ is usually
  ten).
\end{enumerate}

\begin{figure}[htp]
\begin{center}
\epsfig{figure=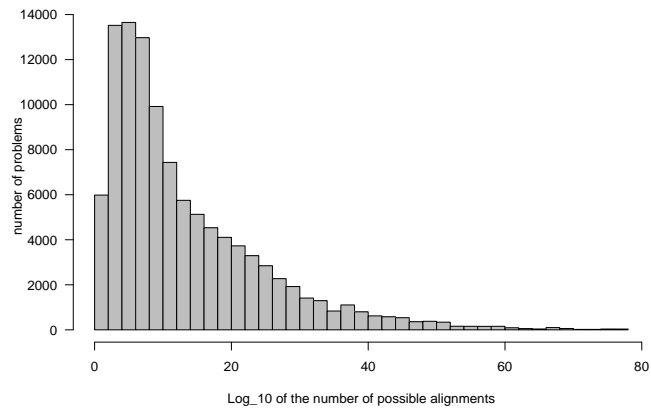, scale=.5}

\end{center}
\caption{Populations of the  3D problems solved during phase $D$ as a
  \emph{$\log_{10}$} function of the size of the search space (number of possible
  alignments).
\label{fig:histrep}}
\end{figure}

Figure~\ref{fig:histrep} shows the distribution of the 
alignment problems needed to be solved during phase $D$ and gives an idea of the amount of computation required by the 3D filter. The number of the problems is about 1,200,000  while 
the size of the largest instance is $6.6~10^{77}$. 

\begin{figure}[htp]
\begin{center}
\epsfig{figure=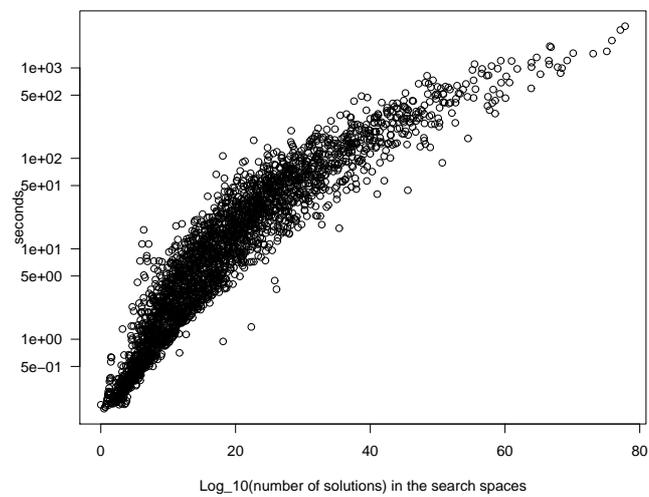, scale=.5}
\end{center}
\caption{Mean CPU time required to solve the 3D problems in phase $D$ as a function 
of their size.
\label{fig:meantime}}
\end{figure}

Figure~\ref{fig:meantime} shows the plot of the mean CPU time required to
solve the 3D problems involved in phase $D$ as a function of the
number of possible alignments\footnote{The mean CPU time here concerns macro-tasks each one containing 
ten (gran3D=10) instances Ali3D  of the same size (see section \ref{subsec:parallel})}.

The purpose of the procedure proposed in the next section is to
distribute all these tasks.

Note that phase $D$ needs to be repeated each time the fitness
functions or the library of templates change, which is a frequent 
case when the program is used in a development phase.

  \subsection{Distribution of the computations: dividing FROST into modules}

The first improvement in the distributed version (DFROST) compared to
the original FROST consists in clearly identifying the different
stages and operations in order to make the entire procedure modular.
The process of computing the scores distributions is dissociated from
the alignment of the query versus the set of templates. We therefore
split the two phases ($D$ and $E$) which used to be interwoven in the
original implementation. Such a decomposition presents several
advantages.

Some of them are:
\begin{itemize}

\item Phase $D$ is completely independent from the query, it can be
  performed as \emph{a preprocessing stage} when it is convenient for
  the program designer.

\item The utilization of the program is \emph{simplified}. Note that
  only the program designer is supposed to execute phase $D$, while
  phase $E$ is executed by an ``ordinary'' user. From a user's
  standpoint DFROST is \emph{significantly faster} than FROST, since
  only phase $E$ is executed  at his request (phase $D$ being
  performed as a preprocessing step).
 
\item The program designer can \emph{easily carry out different
    operations} needed for further developments of the algorithm or
  for database updating such as: adding new filters, changing the
  fitness functions, adding a new template to the library, etc.

\item This organization of DFROST in modules is very \emph{suitable
    for its decomposition in independent tasks} that  can be solved in
  parallel.

\end{itemize}

The latter point is discussed in details in the next section.

  \subsection{Parallel Algorithm}
  \label{subsec:parallel}

We distinguish two kinds of atomic independent tasks in DFROST: the
first is related to solving an instance of a problem of type
\verb|Ali1D|, while the second is associated with solving an instance
of an \verb|Ali3D| problem\footnote{In reality this problem can be
  further decomposed in subtasks. Although non independent, these
  subtasks can be executed in parallel as show in
  \cite{andonov-2004,yanev-2004}. This parallelization could be easily
  integrated in DFROST if necessary.}.

Hence phase $D$ consists in solving 1,200,000 independent tasks of
type \verb|Ali1D| and  \verb|Ali3D|, while phase $E$ consists
in solving several hundreds of independent tasks \verb|Ali1D| and ten
independent tasks \verb|Ali3D|. The final decision requires sorting
and analysis of the $N$ best solutions of type \verb|Ali1D| and the
$N$ best solutions of type \verb|Ali3D|.

There is a couple of important observations to keep in mind in order
to obtain an efficient parallel implementation for DFROST. The first
is that the exact number of tasks is not known in advance. Second,
which is even more important, the tasks are irregular (especially
tasks of type \verb|Ali3D|) with unpredictable and largely
varying execution time. In addition, small tasks need to be aggregated
in macro-tasks in order to reduce data broadcasting overhead. Since
the complexity  of the two types of tasks is different, the granularity
for macro-tasks \verb|Ali1D| should be different from the granularity
for macro-tasks \verb|Ali3D|.

The parallel algorithm that we propose is based on \emph{centralized
  dynamic load balancing}: macro-tasks are dispatched from a
centralized location (pool) in a dynamic way. The work pool is managed
by a ``master'' who gives work \emph{on demand} to idle ``slaves''.
Each slave executes the macro-tasks assigned to it by solving
sequentially the corresponding subproblems (either \verb|Ali1D| or
\verb|Ali3D|). Note that dynamic load balancing is the only reasonable
task-allocation method when dealing with irregular tasks for which the
amount of work is not known prior to execution.

In phase $E$ the pool contains initially several hundreds of tasks of
type \verb|Ali1D|. The master increases the work granularity by
grouping \verb|gran1D| of them in macro-tasks. These macro-tasks are
distributed on demand to the slaves that solve the corresponding
problems. The solutions computed in this way are sent back to the
master and sorted by it locally. The templates associated to the $N$
best scores yield $N$ problems of type \verb|Ali3D|. The master groups
them in batches of size \verb|gran3D| and transmits them to the slaves
where the associated problems are solved. The granularity
\verb|gran1D| is bigger than the \verb|gran3D| granularity. Finally
the slaves send back to the master the computed solutions.

The strategy in phase $D$ is simpler. The master only aggregates tasks
in macro-tasks of size either \verb|gran1D| or \verb|gran3D|, sends
them on demand to idle slaves (where the corresponding problems are
sequentially solved), and finally gathers  the distributions that have
been computed. The master processes the library of templates in a
sequential manner. First, it aims at distributing all  tasks for a
given template to the slaves. However, when the list of tasks for a
given template becomes empty, but the granularity level is not attained,
the master proceeds to distribute tasks from the next template.
This strategy allows to reduce globally the idle time of the
processors.

\newpage
\section{Computational experiments}
\label{sec:experiments}
  \subsection{Running times}
The numerical results presented in this section (see Table
\ref{tab:times}) were obtained on a cluster of 12 Intel(R) Xeon(TM)
CPU 2.4~GHz, 2 Gb Ram, RedHat 9 Linux, connected by a 1~Gb Ethernet
network. The behavior of DFROST was tested by entirely computing the
phase D of the package, i.e. all the distributions for 1125 templates
for both filters.

\begin{table}[htp]
\begin{center}
\begin{tabular}{l|r|r|r|r}
\hline
          & Number of tasks & Wall clock time & Total sequential time & Speed-up\\
\hline
3D filter & 1,104,074      & 3d 3h 20m    & 37d 5h 11m & 11.9 \\
1D filter & 1,107,973      &       31m    &     4h 13m &  8.2 \\
Both filters & 2,202,047   & 3d 3h 51m    & 37d 9h 24m & 11.8 \\
\hline
\end{tabular}
\end{center}
\caption{Comparison of the total time (in days, hours, minutes) taken by a number of 1D and 3D tasks with the corresponding wall clock time after parallelizing the program\label{tab:times}}
\end{table}

In the case of 3D filter, solving 1,104,074 alignments in parallel as
shown on table \ref{tab:times} is very efficient. Comparison of the
total sequential running times with the wall clock time of the master
shows that we obtain a speed-up of about 12, i.e., an efficiency close
to one. In the case of 1D filter, for solving 1,107,973 alignments, the
speed up is lower but then the total sequential time is much shorter
than for solving 3D tasks.

These significant results, obtained on such a large data set, justify
the work done to distribute FROST and prove the efficiency of the
proposed parallel algorithm.

Details from this execution are presented in table \ref{tab:partime}.
The value of the parameters \verb|gran1D| and \verb|gran3D| were experimentally fixed to 1000 and 10 respectively.

\begin{table}[htp]
\begin{footnotesize}
\begin{center}
\begin{tabular}{l|r|r|r|r}
{\bf{Template}}&{\bf{DFROST}}&{\bf{CPU tot}}&{\bf{Cpu av}}&{\bf{NAli}}\\
\hline
\hline
{1BGLA0} & 15455 & 107569 & 113 & 945 \\
1ALO\_0 & 9565 & 96579 & 97 & 995 \\
1CXSA0 & 5988 & 55808 & 58 & 960 \\
1DIK\_0 & 4506 & 46855 & 47 & 977 \\
1BGW\_0 & 4152 & 45286 & 45 & 1000 \\
 1CLC\_0 & 3580 & 37973 & 39 & 969 \\
1AA6\_0 & 3357 & 35819 & 38 & 926 \\
 1DJXB0 & 3025 & 31276 & 31 & 1000 \\
 1DAR\_0 & 2705 & 28671 & 28 & 1000 \\
 1AOZA0 & 2477 & 25156 & 26 & 935 \\
 1AK5\_0 & 2072 & 22326 & 22 & 979 \\
1AUIA0 & 2016 & 22010 & 22 & 1000 \\
 1AOFB0 & 2065 & 21619 & 21 & 1000 \\
 1BHGA0 & 1904 & 20740 & 21 & 980 \\
 1AORA0 & 1920 & 20059 & 20 & 995 \\
1AYL\_0 & 1807 & 18961 & 19 & 973 \\
1EUT\_0 & 1753 & 18883 & 18 & 995 \\
1CTN\_0 & 1535 & 16670 & 16 & 1000 \\
1ECL\_0 & 1439 & 15589 & 16 & 953 \\
 1ATIA0 & 1492 & 15463 & 15 & 980 \\
 1CIY\_0 & 1441 & 15044 & 15 & 1000 \\
 1BYB\_0 & 1307 & 13892 & 14 & 990 \\
 1COY\_0 & 1204 & 13150 & 13 & 957 \\
 1DLC\_0 & 1104 & 11825 & 13 & 907 \\
 1BDP\_0 & 1173 & 12814 & 12 & 995 \\
 1AOP\_0 & 1134 & 12323 & 12 & 1000 \\
 1AG8A0 & 1120 & 12153 & 12 & 990 \\
 1BMFC0 & 1094 & 11338 & 11 & 1000 \\
 1ECFB0 & 1052 & 11254 & 11 & 990 \\
 1DERA0 & 1047 & 11109 & 11 & 1000 \\
 1ALKA0 & 1022 & 10937 & 11 & 965 \\
 1DPE\_0 & 988 & 10626 & 11 & 957 \\
 1DDT\_0 & 973 & 10349 & 10 & 1000 \\
 1AC5\_0 & 907 & 9877 & 9 & 1000 \\
 1CAE\_0 & 913 & 9870 & 9 & 990 \\
 1BMFD0 & 914 & 9467 & 9 & 998 \\
 1DPGA0 & 875 & 9092 & 9 & 1000 \\
 1ASYA0 & 1102 & 8634 & 9 & 952 \\
1LYLA0 & 782 & 8335 & 8 & 990 \\
1BIF\_0 & 657 & 7129 & 7 & 948 \\
1AD3A0 & 629 & 6669 & 6 & 1000 \\
1DNPA0 & 776 & 6580 & 6 & 960
\end{tabular}
\end{center}
\caption{An extract from the execution times (in seconds) when 
    computing the 3D score distributions. The templates for which
    the distributions are calculated are listed in the first column.
    The second column gives the parallel time (the execution time for
    the master) on a cluster of 12 processors. The third column shows
    the CPU sequential time (obtained by adding the CPU times from the
    slaves). The fourth column reports the average CPU time per
    alignment and the last column shows the actual number of sequences
    that have been threaded to calculate the distributions. The value
    of the granularity was fixed to 10.
\label{tab:partime}}
\end{footnotesize}
\end{table}

We can calculate an upper limit for the number of processors beyond
which it is not any more possible to benefit from adding more
processors.  The maximum time for an alignment is 797.4 seconds \ref{tab:stat_analysis}, this
time is the lower limit of the wall clock time for the complete
computation of the distributions for \verb+Ali3d+.  The total CPU time
necessary to calculate all \verb+Ali3D+ alignments is 3,215,460
seconds.  Thus, adding more than \emph{4032 processors}
(3215460/797.4) will not further accelerate the global process.  This
gives a theoretical upper limit. The assumption behind this procedure is that
difficult computations are submitted first.  This strategy was not
implemented in the results presented in \cite{poirriez-2005} since it
requires a criterion for a preliminary running time task estimation.
Our observation on the code behavior when computing all distributions
confirm that a meaningful criterion is the solutions search space (see
figure~\ref{fig:meantime}).  Another criterion could be the observed
in the past running time for a task.

  \subsection{Statistical analysis of the results}
  \label{subsec:stat}

\begin{figure}[htp]
\begin{center}
\epsfig{figure=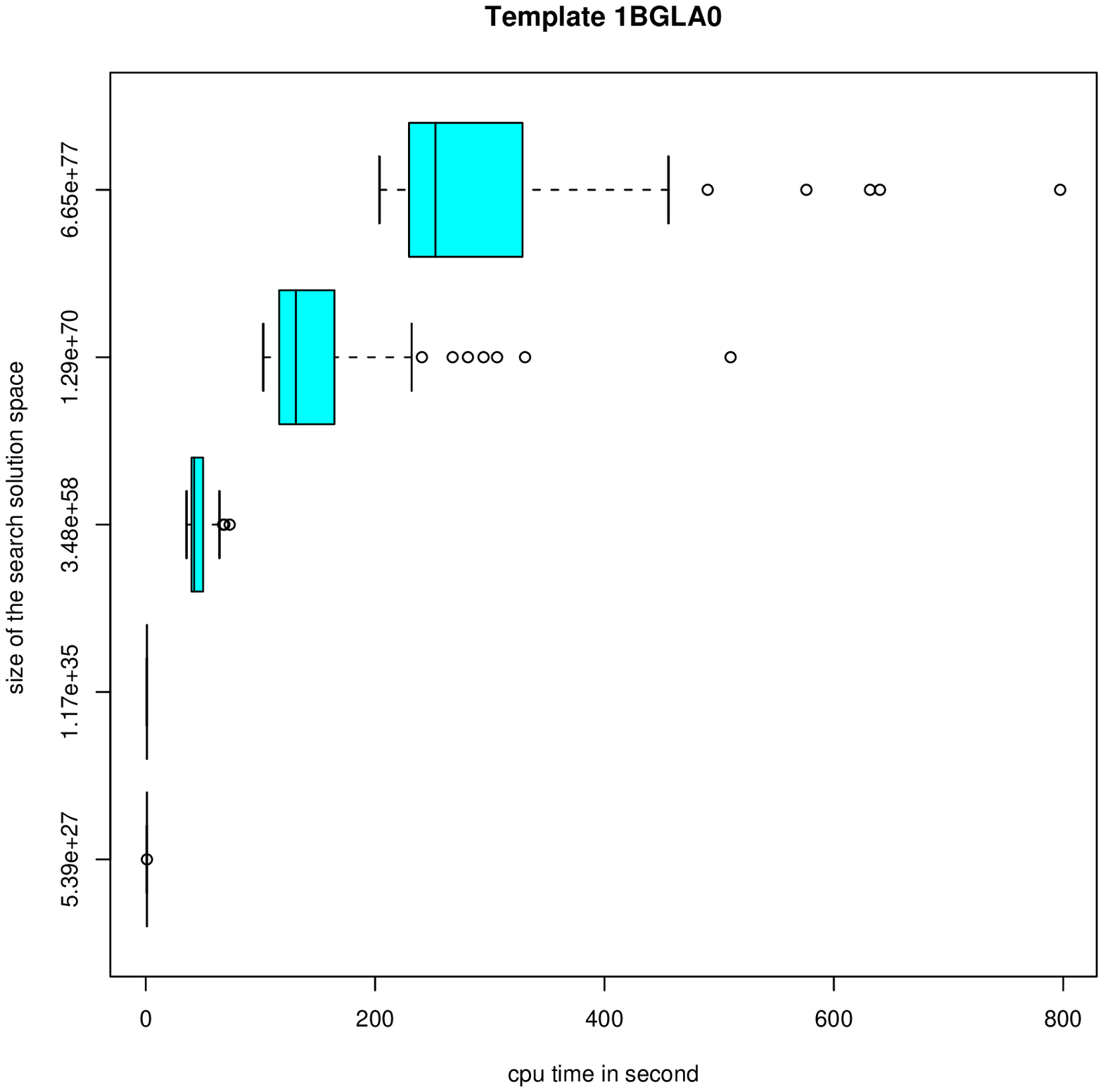, scale=.35}
\epsfig{figure=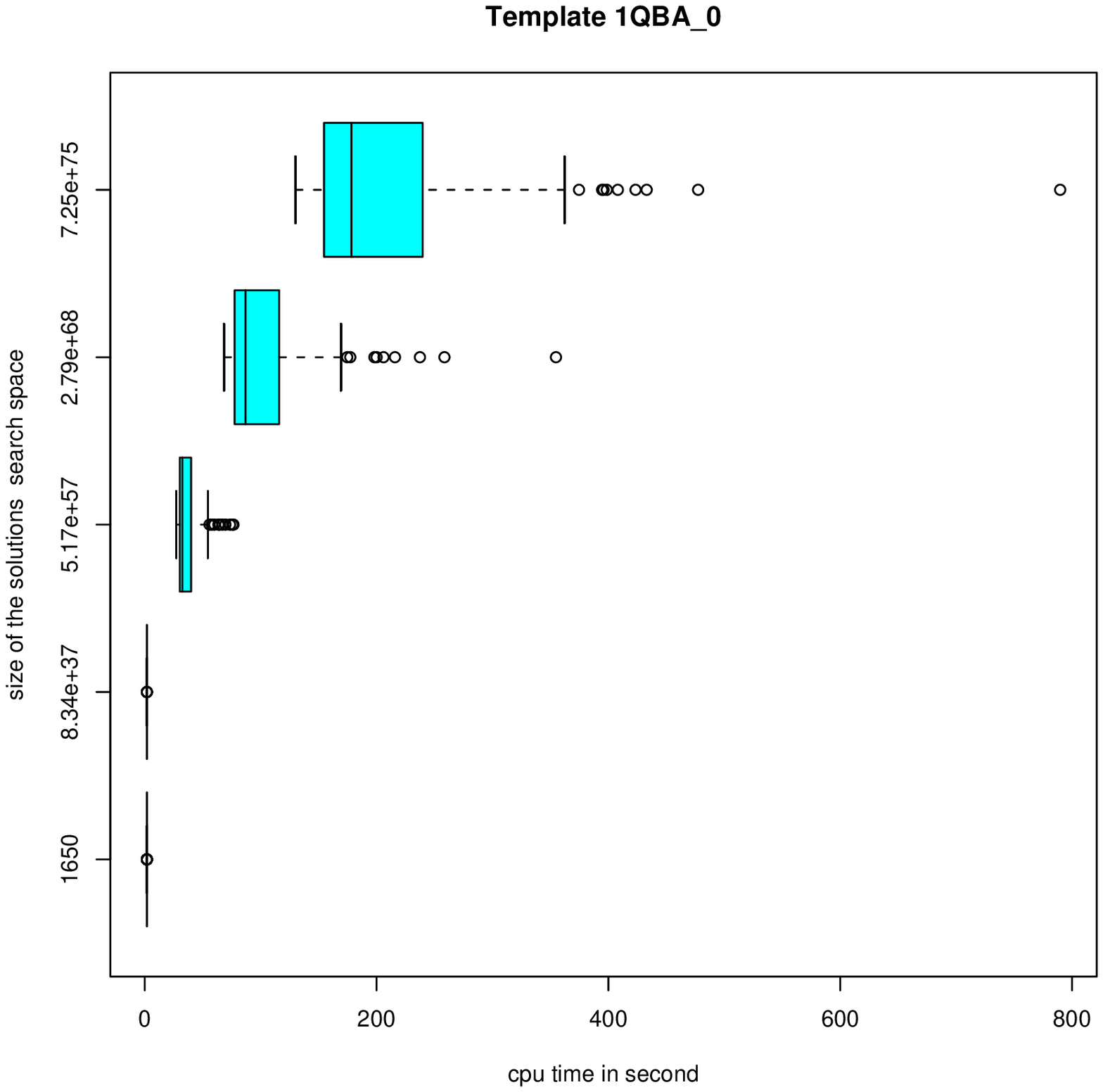, scale=.35}
\end{center}
\caption{Two templates with heavy distribution computations.  1BGLA0
  and 1QBA\_0, are selected from table \ref{tab:stat_analysis} and the
  corresponding box-plots of the distributions running time are plotted
  using the statistical package \textbf{R} \cite{ihaka-1996}. The left
  and right ends of a box correspond to the lower and upper quartiles
  and the middle line corresponds to the median of the distribution.
  Vertical lines, usually called ``whiskers'', go left and right from
  the box to the extreme of the data (here defined as 1.5 times the
  inter-quartile range). Outliers are plotted individually.  Note
that the distribution is not symmetric and exhibits a heavy tail for
longer CPU times.
\label{fig:boxplots}}
\end{figure}

Using this parallel algorithm we were able to compute all 
distributions for the entire FROST templates library. This was never
done with the sequential code, because of large templates like \verb+1BGLA0+ with
sequences as long as 528 amino acids, leading to a number of possible
alignments as large as 6.647E+77. Statistics concerning the running time distribution are 
presented on Figure~\ref{fig:boxplots}. 

On average, the running time distribution \emph{of all}   \verb|Ali1D| 
tasks, is characterized by the following data: 

\begin{center}
\begin{tabular}{ccccc}
minimum & 1st quartile & mean  & 3rd quartile & maximum  \\
\hline
0 s      &   0.03 s      & 0.58 s &  2.32 s      &  797.4 s \\
\end{tabular}
\end{center}

Note that these times correspond to one alignment.

We observed that for 188 templates the computation of the
distributions requires more than one hour CPU time.  Statistical
details concerning the running time of the four most time consuming
templates are presented in table \ref{tab:stat_analysis}.  Remember, that
a PTP instance (i.e. when the query and the 3D structure are fixed) is
considered as an atomic independent task in the current parallel
strategy. Yet, as shown in \cite{andonov-2004,yanev-2004}, such an
instance could be further decomposed in subtasks that could be
executed in parallel. We studied the need for 
implementing this parallelization in the package FROST.  However, taking in account that: 
i) the  number of independent tasks when computing 
distributions is very high; ii) the data from tables \ref{tab:partime}
and \ref{tab:stat_analysis}, as well as their statistical recapitulations
in figure \ref{fig:boxplots}, clearly showing that really hard PTP
instances are rather rare; iii) the speedup reported
in section \ref{sec:experiments} is very satisfactory, we decided,  
for the time being, to stay with the current parallel strategy.

\begin{table}[htp]
\begin{center}
\begin{scriptsize}
\begin{tabular}{|p{0.25cm}|@{\hspace{1mm}}l@{\hspace{1mm}}r|@{\hspace{1mm}}r@{\hspace{1mm}}r@{\hspace{1mm}}r@{\hspace{1mm}}r@{\hspace{1mm}}r@{\hspace{1mm}}r@{\hspace{1mm}}|}
\hline
&\textbf{Nb Sol}&\textbf{NAli}&\textbf{Min}&\textbf{$Q_1$}&\textbf{Med}&\textbf{Mean}&\textbf{$Q_3$}&\textbf{Max}\\
\hline
\multirow{5}{1ex}{\rotatebox{90}{1BGLA0}}&$5.4\;10^{27}$&55& 0.95&  0.96&0.98& 0.97&0.98 & 1.02\\
      &$1.2\;10^{35}$&56& 0.95& 0.96& 0.97& 0.97& 0.98&  1.01\\
      &$3.5\;10^{58}$&192&  35.6&   39.9 &  42.2 &  45.2 &  50.0 &  73.2\\
      &$1.3\;10^{70}$&199&  102.4 &  116.3 &  131.0 &  145.7 &  164.6 &  510.0\\
      &$6.6\;10^{77}$&150&  203.8&   229.7 &  252.6 &  291.7  & 327.5 &  797.4\\
\hline
\multirow{5}{1ex}{\rotatebox{90}{1QBA\_0}}&$1.6\;10^{3}$&58& 1.82&   1.83&   1.83&   1.84&  1.84&   1.89\\
       &$8.3\;10^{37}$&57& 1.82 & 1.83&   1.83& 1.84&   1.84&   1.89\\
       &$5.2\;10^{57}$&197&  27.1&   30.2 &  32.5&   36.3 &  39.8 &  76.6\\
       &$2.8\;10^{68}$&200&  68.4&   77.5&   86.9 & 101.4 & 116.0 & 354.8\\
       &$7.2\;10^{75}$&200&  130.1&   154.7 &  178.3&   207.0 &  239.8  & 789.8\\
\hline
\multirow{5}{1ex}{\rotatebox{90}{1ALO\_0}} &$3.1\;10^{33}$&57& 0.85&  0.87&  0.87&  0.87&  0.88&  0.89\\
        &$6.0\;10^{33}$&57& 0.85&  0.86&  0.87&  0.87&  0.87&  0.89\\
        &$2.5\;10^{57}$& 190&25.8&  29.3&  36.1&  40.8&  46.7& 135.2\\
        &$1.6\;10^{69}$&200& 67.4 &  86.3 & 113.2 & 123.2 & 134.8 & 397.6\\
        &$1.3\;10^{77}$&200&  139.9 & 175.7 & 231.0 & 262.2 & 303.4 & 735.0\\
\hline
\multirow{5}{1ex}{\rotatebox{90}{1YGE\_0}} &$3.4\;10^{23}$&61&0.39&  0.40&  0.41&  0.41&  0.41&  0.43\\
        &$2.8\;10^{45}$&59& 0.40&0.41&0.41&0.41&0.42& 0.42\\
        &$2.1\;10^{55}$&192&  34.8  & 39.9&   43.1 &  47.5 &  48.9 & 139.8\\
        &$6.5\;10^{61}$&173&  71.2  & 80.5 &  89.5&  102.0 & 115.9 & 365.1\\
        &$4.4\;10^{66}$&199&  120.2 &  138.5 &  158.3  & 178.2  & 208.9  & 443.7\\
\hline
\end{tabular}
\end{scriptsize}
\end{center}
\caption{Sequential times \textbf{in seconds} for computing  the 3D score
distributions of four templates selected for their ``difficulty'' (search space
size). For a given template the 5 rows represent alignment of sets of non
related sequences having length respectively equal to: -30\%, -15\%, 0\%, +15\%,
+30\% of the template length.   \textbf{Nb Sol} is the number of possible
alignments that can be generated with  the sequences and the template. This
gives an indication of the difficulty of the problem to solve. \textbf{NAli} is
the number of alignments (sequences) in the corresponding set.  The last six
columns report diverse running time characteristics obtained when aligning the
set of sequences with the corresponding 3D structure: \textbf{Min} is the
minimum value, \textbf{$Q_1$} is the time at the 1st quartile position,
\textbf{Med.} is the time at the median position, \textbf{Mean} is the average
time, \textbf{$Q_3$} is the time at the 3rd quartile position and \textbf{Max}
is the maximum value.
\label{tab:stat_analysis}}
\end{table}

\newpage
\section{Future research directions}
\label{sec:future}

\begin{figure}[htp]
\begin{center}
\epsfig{figure=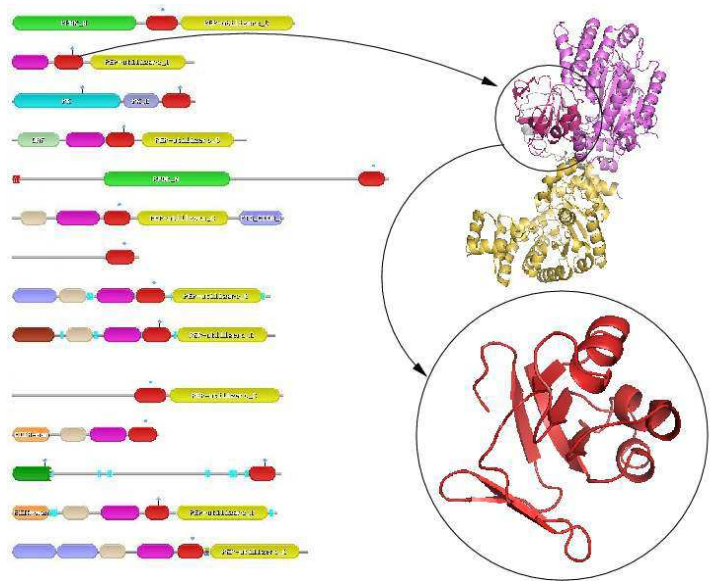, scale=.5}
\end{center}
\caption{Left panel: schematic representation of protein sequences with
  different modules (data from the PFAM database \cite{BCD+04}). In the figure we
  focus on the three modules of the second sequence. These modules are also found in other 
sequences. 
 Upper right panel: the  structure of this sequence (a pyruvate phosphate dikinase) has been solved (PDB
  code 1dik) and the modules have been drawn in similar shades of gray
 in the 3D   structure; lower right panel: zoom on the 3D structure of the
  second module. This module has 102 residues.
  \label{fig:pfam_domains}}
\end{figure}

It is well known that large fractions of the proteins have a modular
organization as shown on Figure \ref{fig:pfam_domains}. Such proteins are
called \emph{multi-domain proteins}. These modules can be detected at
the level of the amino acid sequence as similar subsequences that are
found in different protein sequences. In the 3D structure of the whole
proteins these modules correspond, usually to one, sometimes to
several, substructures called structural domains\footnote{in the
  literature the terms domain and module are often used somewhat
  interchangeably.  In this paper we  restrict the use of module to
  subsequences and domain to 3D substructures} \cite{lesk-1981} (see
the right hand side of Figure \ref{fig:pfam_domains}).

Several cases can occur when studying such multi-domain proteins.  Let
us illustrate this point with the PEP-utilizers domain presented on
Figure \ref{fig:pfam_domains}. 

If one wishes to analyze the PEP-utilizers module family one needs to compare 
 the corresponding sequences over their complete lengths. 
Using global alignment of the sequences (i.e. gaps before the
beginning of a sequence and after its end are penalized) 
will not give a satisfactory result. 
If the goal of the study is to search for the PEP-utilizers module in
a set of sequences (such as those shown in Fig.  \ref{fig:pfam_domains}), 
one must use a semi-global alignment where the gaps
before the beginning,  and after the end of a sequence are
set to zero. This allows the shorter sequence of the PEP-utilizers
module to ``slide'' along the longer sequences until it finds the best match.

The most general case occurs when,  comparing two sequences, for
instance the second and the fifth in Fig. \ref{fig:pfam_domains},
one is trying to analyze  what is common between these sequences.  
This corresponds to carrying out a local alignment, that is, 
finding subsequences in
both sequences that have the maximum score when aligned (for a
given score function).

The local alignment is the most general alignment technique.
Accordingly, this is the convenient  alignment  when
comparing an unknown sequence with a database of sequences, since it is unknown 
beforehand what the  similarity is 
between the query  and the database sequences.

Due to the strong analogy that exists between sequence-sequence 
alignment methods and sequence-structure alignment methods, the above
considerations are also valid, \textit{mutatis mutandis}, for protein
threading methods.

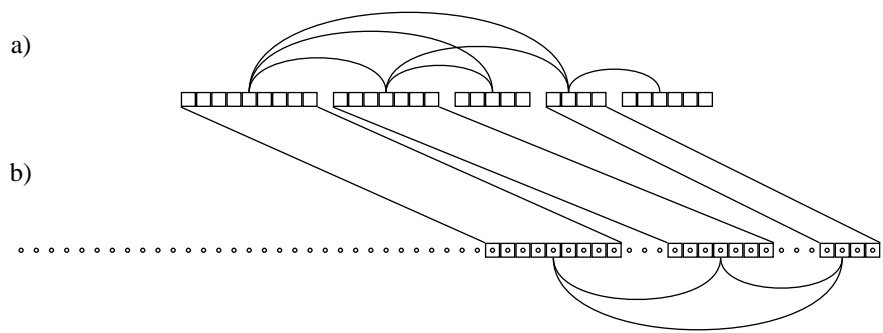
\begin{figure}[htp]
\begin{center}
\begin{pspicture}(7,4)

\psset{linewidth=.5pt,shortput=nab, framesize=0.2, arcangle=90}

\rput(-2.2,3.7){a)}

\fnode(0,3){B11}
\fnode(.2,3){B12}
\fnode(.4,3){B13}
\fnode(.6,3){B14}
\fnode(.8,3){B15}
\fnode(1,3){B16}
\fnode(1.2,3){B17}
\fnode(1.4,3){B18}
\fnode(1.6,3){B19}

\fnode(2,3){B22}
\fnode(2.2,3){B23}
\fnode(2.4,3){B24}
\fnode(2.6,3){B25}
\fnode(2.8,3){B26}
\fnode(3,3){B27}
\fnode(3.2,3){B28}

\fnode(3.6,3){B31}
\fnode(3.8,3){B32}
\fnode(4,3){B33}
\fnode(4.2,3){B34}
\fnode(4.4,3){B35}

\fnode(4.8,3){B41}
\fnode(5,3){B42}
\fnode(5.2,3){B43}
\fnode(5.4,3){B44}

\fnode(5.8,3){B51}
\fnode(6,3){B52}
\fnode(6.2,3){B53}
\fnode(6.4,3){B54}
\fnode(6.6,3){B55}
\fnode(6.8,3){B56}

\ncarc{B15}{B25}
\ncarc{B15}{B33}
\ncarc{B15}{B42}
\ncarc{B25}{B33}
\ncarc{B25}{B42}
\ncarc{B42}{B53}


\rput(-2.2,2){b)}

\cnode(-2.2,1){1pt}{Sm7}
\cnode(-2,1){1pt}{Sm7}
\cnode(-1.8,1){1pt}{Sm7}
\cnode(-1.6,1){1pt}{Sm7}
\cnode(-1.4,1){1pt}{Sm7}
\cnode(-1.2,1){1pt}{Sm7}
\cnode(-1,1){1pt}{Sm7}
\cnode(-.8,1){1pt}{Sm6}
\cnode(-.6,1){1pt}{Sm5}
\cnode(-.4,1){1pt}{Sm4}
\cnode(-.2,1){1pt}{Sm3}
\cnode(0,1){1pt}{Sm2}
\cnode(.2,1){1pt}{Sm1}
\cnode(.4,1){1pt}{S1}
\cnode(.6,1){1pt}{S2}
\cnode(.8,1){1pt}{S3}
\cnode(1,1){1pt}{S4}
\cnode(1.2,1){1pt}{S5}
\cnode(1.4,1){1pt}{S6}
\cnode(1.6,1){1pt}{S7}
\cnode(1.8,1){1pt}{S8}
\cnode(2,1){1pt}{S9}
\cnode(2.2,1){1pt}{S10}
\cnode(2.4,1){1pt}{S11}
\cnode(2.6,1){1pt}{S12}
\cnode(2.8,1){1pt}{S13}
\cnode(3,1){1pt}{S14}
\cnode(3.2,1){1pt}{S15}
\cnode(3.4,1){1pt}{S16}
\cnode(3.6,1){1pt}{S17}
\cnode(3.8,1){1pt}{S18}
\cnode(4,1){1pt}{S19}
\cnode(4.2,1){1pt}{Sm1}
\cnode(4.4,1){1pt}{S1}
\cnode(4.6,1){1pt}{S2}
\cnode(4.8,1){1pt}{S3}
\cnode(5,1){1pt}{S4}
\cnode(5.2,1){1pt}{S5}
\cnode(5.4,1){1pt}{S6}
\cnode(5.6,1){1pt}{S7}
\cnode(5.8,1){1pt}{S8}
\cnode(6,1){1pt}{S9}
\cnode(6.2,1){1pt}{S10}
\cnode(6.4,1){1pt}{S11}
\cnode(6.6,1){1pt}{S12}
\cnode(6.8,1){1pt}{S13}
\cnode(7,1){1pt}{S14}
\cnode(7.2,1){1pt}{Sm1}
\cnode(7.4,1){1pt}{S1}
\cnode(7.6,1){1pt}{S2}
\cnode(7.8,1){1pt}{S3}
\cnode(8,1){1pt}{S4}
\cnode(8.2,1){1pt}{S5}
\cnode(8.4,1){1pt}{S6}
\cnode(8.6,1){1pt}{S7}
\cnode(8.8,1){1pt}{S8}
\cnode(9,1){1pt}{S9}
\cnode(9.2,1){1pt}{S10}

\fnode(4,1){D11}
\fnode(4.2,1){D12}
\fnode(4.4,1){D13}
\fnode(4.6,1){D14}
\fnode(4.8,1){D15}
\fnode(5,1){D16}
\fnode(5.2,1){D17}
\fnode(5.4,1){D18}
\fnode(5.6,1){D19}

\fnode(6.4,1){D22}
\fnode(6.6,1){D23}
\fnode(6.8,1){D24}
\fnode(7,1){D25}
\fnode(7.2,1){D26}
\fnode(7.4,1){D27}
\fnode(7.6,1){D28}

\fnode(8.4,1){D41}
\fnode(8.6,1){D42}
\fnode(8.8,1){D43}
\fnode(9,1){D44}

\ncarc[arcangle=270]{D15}{D25}
\ncarc[arcangle=270]{D15}{D42}
\ncarc[arcangle=270]{D25}{D42}


\psline(-0.1,2.9)(3.9,1.1)
\psline(1.7,2.9)(5.7,1.1)
\psline(1.9,2.9)(6.3,1.1)
\psline(3.3,2.9)(7.7,1.1)
\psline(4.7,2.9)(8.3,1.1)
\psline(5.5,2.9)(9.1,1.1)
\end{pspicture}
\end{center}
\caption{Local alignment. a) A template containing five blocks. b) A sequence of 58 amino acids.
 On its right-hand site this sequence contains a structural domain which exhibits a good similarity
 to the template when only three blocks are aligned. To obtain this optimal 
alignment (i.e. giving the best score), two blocks have to be  omitted.}
\label{fig:flex_align}
\end{figure}

In section \ref{subsec:signif} we mentioned that FROST permits  \emph{only}
global alignment of a sequence with a core. Even more, to the best of
our knowledge, no current protein threading approach exists, that  uses 
non-local score functions for providing an exact solution, 
and that is able to carry out semi-global and local alignments. 
Some ideas to tackle this problem have been presented by G. Collet and al.  in
\cite{collet-2006,collet-2006b} where mathematical formulations, based on MIP
models for semi-global and local sequence/structure alignment,  are discussed. 
The latest one is also called \emph{flexible alignment} since
it allows  omissions of blocks during the alignment process (see Fig.
\ref{fig:flex_align}).

Semi-global and flexible alignments raise a number of new questions.
Performing such alignments necessitates the alignment of cores with
potentially very long sequences (the largest proteins known are up to
10\,000 residues long).  The process of computing distributions  (see \ref{subsec:signif}) 
needs to be significantly modified in the context of arbitrarily long sequences. 
In addition, these types of alignment will 
drastically increase the solution space and the corresponding
running  time.  In order to  manage such an increase of the
computational requirements the future semi-global and
flexible alignment algorithms will need  more and more parallel and distributed  
computing.

\newpage
\section{Conclusion}
\label{sec:conclude}

Fold recognition (protein threading) is rather typical of problems
that occur in bioinformatics. It requires knowledge from different
disciplines: biology for the definition of cores, physical-chemistry
for the development of score functions, computer science for the
conception of efficient alignment algorithms and statistics for
explaining the significance of the alignment score.

Sequence comparison methods play an outstanding role for exploiting
protein sequence data, in particular for {\it in silico} functional
analysis. These methods are versatile and extremely efficient as long
as close homologs are considered. Fold recognition techniques are
intended to replace them when the much more difficult case of remote
homologs needs to be tackled. Unfortunately, fold recognition
techniques are computer intensive and, for the moment, are less
universal. In particular the problem of fold recognition has received
a satisfactory solution only for the case of global alignments whereas,
due to the protein modularity properties, semi-global and local
alignments are urgently needed. Fold recognition methods are also
plagued by the lack of a statistical theory permitting to assess the
significance of alignment scores. Our goal, in the near future, is to
set fold recognition methods on an equal footing with sequence
alignment methods in terms of available types of alignment and
assessment of the alignment score significance.

Of course, due to the inescapable NP-hard property of fold recognition
alignment algorithms, these methods will always be more demanding in
terms of computer resources than sequence alignments, although we are
able to achieve pruning peak rate as high as $10^{74}$ per second for
global alignments. However, as shown in this paper, it is possible to
harness the power of grid computing to perform the heavy calculations
that will be needed to analyze the 500 currently sequenced microbial
genomes and the further thousand that are to be released next year.

\newpage
\bibliography{bibli}

\begin{thebibliography}{10}

\bibitem{AMS+97}
SF~{Altschul}, TL~{Madden}, AA~{Schaffer}, J~{Zhang}, Z~{Zhang}, W~{Miller},
  and DJ~{Lipman}.
\newblock Gapped blast and psi-blast: a new generation of protein database
  searchprograms.
\newblock {\em Nucleic Acids Res}, 25:3389--402, 1997.

\bibitem{brenner-1998}
S.E. Brenner, C.~Chothia, and T.J. Hubbard.
\newblock Assessing sequence comparison methods with reliable structurally
  identified distant evolutionary relationships.
\newblock {\em Proc Natl Acad Sci U S A}, 95:6073--6078, 1998.

\bibitem{chothia-2004}
C.~Chothia.
\newblock One thousand families for the molecular biologist.
\newblock {\em Nature Biotechnology}, 22:1317--1321, 2004.

\bibitem{OJT94}
CA~{Orengo}, DT~{Jones}, and JM~{Thornton}.
\newblock Protein superfamilies and domain superfolds.
\newblock {\em Nature}, 372:631--4, 1994.

\bibitem{pearl-2003}
F.M. Pearl, C.F. Bennett, J.E. Bray, A.P. Harrison, N.~Martin, A.~Shepherd,
  I.~Sillitoe, J.~Thornton, and C.A. Orengo.
\newblock The cath database: an extended protein family resource for structural
  and functional genomics.
\newblock {\em Nucleic Acids Research}, 31(1):452--455, 2003.

\bibitem{andreeva-2004}
A.~Andreeva, D.~Howorth, S.E. Brenner, T.J.P. Hubbard, C.~Chothia, and A.G.
  Murzin.
\newblock Scop database in 2004: refinements integrate structure and sequence
  family data.
\newblock {\em Nucleic Acids Research}, 32:226--229, 2004.

\bibitem{marin-2002}
A.~Marin, J.Pothier, K.~Zimmermann, and J-F. Gibrat.
\newblock Frost: A filter based fold recognition method.
\newblock {\em Proteins}, 49(4):493--509, 2002.

\bibitem{lathrop-1994}
R.H. Lathrop.
\newblock The protein threading problem with sequence amino acid interaction
  preferences is {NP}-complete.
\newblock {\em Protein Engineering}, 255:1059--1068, 1994.

\bibitem{lathrop-1996}
R.H. Lathrop and T.F. Smith.
\newblock Global optimum protein threading with gapped alignment and empirical
  pair potentials.
\newblock {\em Journal of Molecular Biology}, 255:641--665, 1996.

\bibitem{yanev-2004}
N.~Yanev and R.~Andonov.
\newblock Parallel divide\&conquer approach for the protein threading problem.
\newblock {\em Concurrency and Computation: Practice and Experience},
  16:961--974, 2004.

\bibitem{andonov-2004}
R.~Andonov, S.~Balev, and N.~Yanev.
\newblock Protein threading problem: From mathematical models to parallel
  implementations.
\newblock {\em INFORMS Journal on Computing}, 16(4):393--405, 2004.
\newblock Special Issue on Computational Molecular Biology/Bioinformatics, Eds.
  H. Greenberg, D. Gusfield, Y. Xu, W. Hart, M. Vingro.

\bibitem{xu-2003}
J.~Xu, M.~Li, G.~Lin, D.~Kim, and Y.~Xu.
\newblock Raptor: optimal protein threading by linear programming.
\newblock {\em Journal of Bioinformatics and Computational Biology},
  1(1):95--118, 2003.

\bibitem{xu-2000}
Y.~Xu and D.~Xu.
\newblock Protein threading using prospect: design and evaluation.
\newblock {\em Proteins}, 40(3):343--354, 2000.

\bibitem{balev-2004}
Stefan Balev.
\newblock Solving the protein threading problem by lagrangian relaxation.
\newblock In Jonassen and J.~Kim, editors, {\em 4th International Workshop on
  Algorithms in Bioinformatics, Bergen, Norway. Volume 3240 of LNCS/LNBI WABI
  2004}, pages 182--193, 2004.

\bibitem{li-2003}
W.W. Li, R.W. Byrnes, J.~Hayes, V.M. Reyes, A.~Birnbaum, A.~Shahab, C.~Mosley,
  D.~Pekurovsky, G.B. Quinn, I.N. Shindyalov, H.~Casanova, L.~Ang, F.~Berman,
  M.A. Miller, and P.E. Bourne.
\newblock The encyclopedia of life project: Grid software and deployment.
\newblock {\em Special Issue on Grid Systems for Life Sciences. New Generation
  Computing}, 2003.

\bibitem{steinke-2003}
Thomas Steinke.
\newblock Alignment \& threading on massively parallel computers.
\newblock Technical report, Berlin Center for Genome Based Bioinformatics,
  2003.

\bibitem{xu-1998}
Y.~Xu, D.~Xu, and E.C. Uberbacher.
\newblock An efficient computational method for globally optimal threading.
\newblock {\em Journal of Computational Biology}, 5:597--614, 1998.

\bibitem{Altschul91}
SF~{Altschul}.
\newblock Amino acid substitution matrices from an information theoretic
  perspective.
\newblock {\em J Mol Biol}, 219:555--65, 1991.

\bibitem{DSO78}
MO~{Dayhoff}, RM~{Schwartz}, and BC~Orcutt.
\newblock {\em Atlas of protein sequence and structure}, volume~5, chapter A
  model of evolutionary change in proteins, pages 345--352.
\newblock National Biomedical Research Foundation, Washington DC, 1978.

\bibitem{HH92}
S~{Henikoff} and JG~{Henikoff}.
\newblock Amino acid substitution matrices from protein blocks.
\newblock {\em Proc Natl Acad Sci U S A}, 89:10915--9, 1992.

\bibitem{NW70}
SB~{Needleman} and CD~{Wunsch}.
\newblock A general method applicable to the search for similarities in the
  aminoacid sequence of two proteins.
\newblock {\em J Mol Biol}, 48, 1970.

\bibitem{SW81}
TF~{Smith} and MS~{Waterman}.
\newblock Identification of common molecular subsequences.
\newblock {\em J Mol Biol}, 147:195--7, 1981.

\bibitem{Pearson00}
WR~{Pearson}.
\newblock Flexible sequence similarity searching with the fasta3 program
  package.
\newblock {\em Methods Mol Biol}, 132:185--219, 2000.

\bibitem{BCH98}
SE~{Brenner}, C~{Chothia}, and TJ~{Hubbard}.
\newblock Assessing sequence comparison methods with reliable
  structurallyidentified distant evolutionary relationships.
\newblock {\em Proc Natl Acad Sci U S A}, 95:6073--8, 1998.

\bibitem{MGB95}
T~{Madej}, JF~{Gibrat}, and SH~{Bryant}.
\newblock Threading a database of protein cores.
\newblock {\em Proteins}, 23:356--69, 1995.

\bibitem{akutsu-1999}
T.~Akutsu and S.~Miyano.
\newblock On the approximation of protein threading.
\newblock {\em Theoretical Computer Science}, 210:261--275, 1999.

\bibitem{yanev-2003}
Nicola Yanev and Rumen Andonov.
\newblock Solving the protein threading problem in parallel.
\newblock In {\em IPDPS '03: Proceedings of the 17th International Symposium on
  Parallel and Distributed Processing}, page 157.1. IEEE Computer Society,
  2003.

\bibitem{xu-2003-PSB}
J.~Xu, M.~Li, G.~Lin, D.~Kim, and Y.~Xu.
\newblock Protein structure prediction by linear programming.
\newblock In {\em Proceedings of The 7th Pacific Symposium on Biocomputing
  (PSB)}, pages 264--275, 2003.

\bibitem{lancia-2004}
G.~Lancia.
\newblock Integer programming models for computational biology problems.
\newblock {\em J. Comput. Sci. \& Technol.}, 19(1):60--77, 2004.

\bibitem{Blazewicz-2006}
J.~Blazewicz, P.~Lukasiak, and M.~Milostan.
\newblock Some operations research methods for analyzing protein sequences and
  structures.
\newblock {\em 4OR A Quarterly Journal of Operations Research}, 4(2):91--123,
  2006.

\bibitem{KA90}
S~{Karlin} and SF~{Altschul}.
\newblock Methods for assessing the statistical significance of molecular
  sequence features by using general scoring schemes.
\newblock {\em Proc Natl Acad Sci U S A}, 87:2264--8, 1990.

\bibitem{MFS00}
LA~{Mirny}, AV~{Finkelstein}, and EI~{Shakhnovich}.
\newblock Statistical significance of protein structure prediction by
  threading.
\newblock {\em Proc Natl Acad Sci U S A}, 97:9978--83, 2000.

\bibitem{marin-2002b}
K.~Zimmermann A.~Marin, J.Pothier and J-F. Gibrat.
\newblock {\em Protein structure prediction: bioinformatic approach}, chapter
  Protein threading statistics: an attempt to assess the significance of a fold
  assignment to a sequence.
\newblock International University line, 2002.

\bibitem{setubal-1997}
J.~Setubal and J.~Meidanis.
\newblock {\em Introduction to computational molecular biology}.
\newblock PWS publishing company, 1997.

\bibitem{lathrop-1998}
R.H. Lathrop, R.G.~Rogers Jr., J.~Bienkowska, B.K.M. Bryant, L.J. Buturovic,
  C.~Gaitatzes, R.~Nambudripad, J.V. White, and T.F. Smith.
\newblock {\em Computational Methods in Molecular Biology}, chapter~12, pages
  227--283.
\newblock Elsevier Science, 1998.

\bibitem{yanev-2006}
N.~Yanev, P.~Veber, R.~Andonov, and S.~Balev.
\newblock Lagrangian approaches for a class of matching problems in
  computational biology.
\newblock Rapport de recherche RR-5973, INRIA, August 2006.
\newblock to appear in Computers and Mathematics with Applications, special
  issue on Computational Biology, Edt.Roberto Tadei.

\bibitem{veber-2005}
P.~Veber, N.~Yanev, R.~Andonov, and V.~Poirriez.
\newblock Optimal protein threading by cost-splitting.
\newblock In {\em WABI'05 (5th Workshop on Algorithms in Bioinformatics)},
  volume 3692 of {\em Lecture Notes in Computer Science}, pages 365--375.
  Springer, 2005.

\bibitem{poirriez-2005}
V.~Poirriez, R.~Andonov, A.~Marin, and J-F. Gibrat.
\newblock Frost: Revisited and distributed.
\newblock In {\em IPDPS '05: Proceedings of the 19th IEEE International
  Parallel and Distributed Processing Symposium (IPDPS'05) - Workshop 7}, page
  200.1, Washington, DC, USA, 2005. IEEE Computer Society.

\bibitem{ihaka-1996}
Ross Ihaka and Robert Gentleman.
\newblock R: A language for data analysis and graphics.
\newblock {\em Journal of Computational and Graphical Statistics},
  5(3):299--314, 1996.

\bibitem{lesk-1981}
A.M.Lesk and G.D. Rose.
\newblock Folding units in globular proteins.
\newblock {\em PNAS}, 78:4304--4308, 1981.

\bibitem{BCD+04}
A.~Bateman, L.~Coin, R.~Durbin, R.D. Finn, V.~Hollich, Jones Griffiths,
  A.~Khanna, M.~Marshall, S.~Moxon, E.L. Sonnhammer, D.J. Studholme, C.~Yeats,
  and S.R. Eddy.
\newblock The {Pfam} protein families database.
\newblock {\em Nucleic Acids Res}, 32:D138--41, 2004.

\bibitem{collet-2006}
G.~Collet, A.~Marin, N.~Yanev, R.~Andonov, and J-F. Gibrat.
\newblock Implementing a semi-global alignment algorithm for protein threading
  methods that use non-local score functions.
\newblock In {\em Poster of the ROADEF conference}, 2006.
\newblock in French.

\bibitem{collet-2006b}
G.~Collet, N.~Yanev, A.~Marin, R.~Andonov, and J-F. Gibrat.
\newblock A flexible model for protein fold recognition.
\newblock In A.~Denise, P.~Durrens, S.~Robin, E.~Rocha, A.~de~Daruvar, and
  A.~Groppi, editors, {\em Septi\`{e}mes Journées Ouvertes de Biologie,
  Informatique et Math\'{e}matiques (JOBIM)}, pages 215--216, 2006.

\end{thebibliography}

\newpage
\tableofcontents

\end{document}